# ProTEUS: A mutation-resolved world model for closed-loop protein evolution

Yikun Zhang[1,5#,6#,*], Minghao Guo[2,*], Guangxue Xu[3], Peter Yichen Chen[4], Wojciech Matusik[2], Wengong Jin[5,6], Jingjing Chai[1], Kaiyi Jiang[1,7,†]

[1] Foundry Bioscience Inc.

Burlingame, CA, 94010, United States of America

[2] Department of Computer Science

Massachusetts Institute of Technology

Cambridge, MA, 02139, United States of America

[3] Department of Pathology, Department of Genetics,

Stanford University School of Medicine

California, 94305, United States of America

[4] Department of Computer Science

The University of British Columbia

Vancouver, BC Canada V6T 1Z4

[5] Khoury College of Computer Sciences

Northeastern University, Boston, MA, 02115, United States of America

[6] Broad Institute of MIT and Harvard

Cambridge, MA, 02142, United States of America

[7] Omenn-Darling Bioengineering Institute

Princeton University

Princeton, NJ, 08540, United States of America

[*]These authors contributed equally

[#]Present address

†Correspondence to: **kaiyi.jiang@princeton.edu**

# Abstract

Protein engineering requires navigating rugged fitness landscapes, where small sequence changes can produce large functional differences, but experimental screening can only sample a small fraction of sequence space. Thus, a central challenge is to efficiently traverse the large combinatorial space of mutations in silico. We introduce ProTEUS, a mutation-resolved world model for closed-loop protein evolution. Pretrained on deep mutational scanning assays, ProTEUS learns a joint representation of sequence, evolutionary, and structural information using MutationFormer, a mutation-biased attention architecture. Variants are represented as states in this joint latent space, and mutations define actions whose functional consequences are predicted. To support exploration of many mutants, we developed a fast, physics-constrained all-atom predictor that utilizes explicit geometric constraint enforcement to achieve >10 fold increase in speed than AlphaFold3 with guaranteed physical validity. Lastly, ProTEUS refines its prediction and understanding of protein activity through closed loop experimental feedback to guide evolution toward a desired phenotype. ProTEUS surpasses established models in both zero-shot and supervised fine-tuning mutation prediction tasks. We experimentally demonstrate its utility in two closed-loop multi-objective engineering campaigns on aldehyde dehydrogenase and levansucrase, yielding variants with >10 fold higher activity than wild type. ProTEUS provides a world-model framework that connects virtual evaluation of mutational outcomes with experimental learning for efficient protein evolution.

# Introduction

Protein engineering expands the capabilities of biotechnology and medicine by improving a desired activity beyond what is offered in nature including catalytic activity, binding affinity, stability and expression (*1*–*3*). Directed evolution has played a central role in this pursuit, which achieves these goals through repeated mutation and selection, but experimental screening, including continuous directed evolution techniques, can only sample a small fraction of the full sequence space (*4*, *5*). Although deep mutational scanning for all single mutants of a protein is possible, epistasis still complicates the route to find the best sequence which may process multiple mutations from wild-type (WT) (*6*–*8*). A central challenge is thus to build a computational model to evaluate a large number of mutations in silico to guide selection of sequences for experimental testing and then use experimental feedback to improve activity predictions for successive rounds of variant selection toward a desired phenotype (*3*, *9*, *10*).

Various models have been deployed to score mutations in a zero-shot setting to reduce the initial experimental workload in a protein engineering campaign before target specific measurements are obtained. Sequence-based models, including the ESM family, learn from natural proteins and use amino acid likelihoods to estimate substitution effects (*11*–*14*). MSA Transformer, Tranception, PoET and Protriever incorporate family-specific evolutionary information (*15*–*18*), whereas inverse folding models like ESM-IF and ProteinMPNN assess sequence compatibility with a specified protein backbone (*19*, *20*). Multimodal models such as ESM-3, SaProt and ProSST combine sequence and structural information (*21*–*23*) and local geometry, residue contacts and molecular surfaces could provide further context for enhanced prediction (*24*–*26*). Lastly, retrieval-augmented models can incorporate homologous sequences and evolutionary information to help with mutational effect prediction (*27*, *28*). Together, these approaches support virtual screening through estimates of evolutionary plausibility and structural compatibility, but so far still exhibit only a moderate level of correlation with experimentally measured protein activity especially for proteins that are less represented in evolution (*3*, *29*).

The moderate performance of these zero-shot predictions can be partly attributed to the misalignment between evolutionary selection and a particular activity being optimized. Natural sequence distributions reflect selection across biological constraints for a particular organism, which can differ significantly from the activity targeted in an experiment (*30*). Favorable sequence, structural or evolutionary scores therefore do not guarantee higher protein activity (*1*, *3*, *30*). To circumvent this problem, functional measurements can be integrated to adapt the model to an engineering objective (*31*). In this pursuit, EVOLVEpro uses protein language model representations to guide iterative evolution from small experimental datasets (*3*). MULTI-evolve uses a similar approach on higher order mutants (*1*). However, these two models still suffer from low hit rates in selecting a high-activity initial set of mutations for experimental characterization and poor generalization to untested mutants, especially higher-order mutants, during active learning.

In light of these challenges, we developed ProTEUS, a world model for closed-loop **PROT**ein **E**volution with high m**U**tation re**S**olution (Fig. 1A-C). Inside ProTEUS, protein variants are represented as states in a joint embedding space of sequence, evolutionary and structural information. We pretrained this joint latent representation on 182 deep mutational scanning (DMS) assays spanning 161 protein backgrounds with diverse protein properties including stability, organismal fitness, activity, expression and binding (*29*). At its core, MutationFormer aligns sequence, structural, and multiple sequence alignment (MSA) features residue by residue and integrates them through attention biased toward mutations (Fig. 1D). This design focuses the model's attention on the mutations defining a candidate while allowing their predicted effects to depend on the surrounding protein context. To supply physically valid structural inputs for large libraries of higher-order candidates, we developed a fast, physics-constrained all-atom predictor. The predictor combines two denoising steps with differentiable Gauss–Seidel projection to enforce geometric and stereochemical constraints during training and inference (*32*). The resulting structures provide candidate-specific spatial context for the joint representation, supporting virtual evaluation of proposed mutations before laboratory testing. During pretraining, ProTEUS exhibited data-scaling behavior with respect to protein activity data (Fig. 2C), as seen in the training of large language models (LLM) and protein language models (PLM) where increase in training input data or model size leads to better performance and generalization. This reveals a potentially new axis of scaling for protein engineering foundation models to capitalize on more protein activity data of various forms.

ProTEUS guides successive rounds of evolution through a nomination strategy linked to experimental feedback (Fig. 1C). Zero-shot scores guide the initial selection of high-fitness variants for experimental testing. Measurements from each round update activity predictions, and the nomination strategy combines these predictions with accumulated experimental evidence to select subsequent higher order candidates. Computational benchmarks show improved mutation-effect prediction in both zero-shot and supervised fine-tuning settings relative to the evaluated baselines. ProTEUS also improves higher-order fitness prediction and recovery of high-fitness variants during simulated evolution compared to EVOLVEpro and MULTI-evolve. In laboratory evolution of levansucrase and aldehyde dehydrogenase, ProTEUS can find variants with >10 fold higher activity than the starting wild-type protein within an experimental budget of 100 sequences. Together, these results establish a world-model framework that links predicted mutational consequences to experimental observations, enabling experimental learning to guide protein evolution toward desired functions.

# Results

### Mutation-resolved learning improves protein fitness prediction

For scoring mutational effect on protein activity, where most of the benchmarks and datasets center on single mutants (*29*, *33*), evaluation requires distinguishing candidate states that share most of their sequence but differ substantially in function. Conventionally, pooling residue-level features into a single protein representation is used but this can obscure local information relevant to prediction (*13*, *34*, *35*). The challenge thus is to focus the attention on the few substituted residues while preserving the surrounding context that determines their effects. Selective attention and gating can emphasize information relevant to a prediction (*36*–*39*), while controlling anomalous activations can reduce attention and feature-map artifacts (*40*). These principles motivate us to build a mutation-resolved representation in which mutations receive explicit attention and are interpreted jointly with their sequence, evolutionary and structural context (Fig. 1D). In particular, amino acid identity, MSA and structural features were aligned residue by residue, and an explicit mutation mask marked differences from the wild-type (WT) sequence. MutationFormer integrated these features through a 12-layer transformer with attention biased toward substituted positions, while retaining the surrounding protein as context.

To establish this joint embedding space, we pretrained ProTEUS on 182 substitution DMS assays from ProteinGym (*29*) to predict measured variant fitness from multimodal inputs (Fig. 1E, tables S1-2). The pretraining data includes experimentally measured stability, organismal fitness, activity, expression and binding across human, prokaryotic, other eukaryotic and viral proteins (Fig. 2A). We screened 217 ProteinGym assays against 38 reference sequences from held-out datasets reserved for downstream evaluation to test model generalization. After excluding assays that matched these evaluation targets or met the sequence-homology exclusion criteria, we retained 182 assays for pretraining. We characterized sequence and structural level of homology to make sure there is no aberrant leakage and found no hits with ≥ 30% sequence-identity that are either above 50% query-coverage thresholds or query-normalized TM-scores ≥0.5 for structural threshold. (fig. S1). Within this pretraining data, ProTEUS achieved a Spearman correlation of 0.933 and a mean squared error (MSE) of 0.115 on the test set. To compare different model architectures' ability to learn this mutational dataset, we adopted a similar training procedure and found that PoET-2, the strongest evaluated baseline by rank correlation in this comparison, only achieved 0.861, showing ProTEUS's ability to use a joint latent space to efficiently learn mutational effect on protein activity (*41*) (Fig. 2B).

We next quantified how the world model's predictive accuracy scaled with functional pretraining data. With the architecture and evaluation sets held fixed, increasing training data 80-fold reduced test MSE from 0.262 to 0.115, a 56.3% reduction (Fig. 2C). Across the measured range, MSE followed an approximate power-law relationship with training-data size. This fit corresponded to an approximately 12.1% reduction in MSE for each doubling of training data. Spearman correlation of the held out test set increased concurrently from 0.842 to 0.933, showing that additional supervision improved both numerical fitness prediction and variant ranking. Although absolute gains diminished as the dataset expanded, both metrics continued to improve at the largest training fraction. These results identify experimental activity measurements as a potential scalable source of supervision for refining the world model's predictions of untested variants.

Next, we used ablations to assess the contributions of mutation-biased attention and multimodal context to fitness prediction (Fig. 2D). Removing mutation bias from an otherwise matched model reduced Spearman correlation from 0.840 to 0.606; a bias strength of 10 performed best among the values tested. In the input-modality comparison, sequence representations alone achieved a correlation of 0.807, whereas joint sequence, MSA and structural representations achieved 0.938. Redundancy filtering and weighted sampling of MSAs produced similar performance (Fig. 2D). Together, these comparisons supported both explicit attention to substituted residues and multimodal integration of a protein's evolutionary and structural context for optimal understanding of mutational effect.

**ProTEUS predicts mutation effects and guides iterative variant selection**

Virtual exploration of a new protein requires predicting mutational outcomes before target-specific functional measurements are available. We therefore tested whether mutation–fitness relationships learned across the pretraining datasets enable generalization to proteins excluded from pretraining (fig. S1). Across 19 held-out DMS assays containing single- and multiple-substitution variants, ProTEUS achieved the highest macro-averaged zero-shot Spearman correlation compared to other models (Fig. 2E; fig. S2A).

We then tested whether zero-shot predictions supported the selection of high-fitness variants, since most protein engineering campaigns rely on an initial set of high activity single mutants to build combinatorial libraries. In this pursuit, MULTI-evolve previously reported a method by taking an ensemble of four deep learning models' predictions with high accuracy (*1*). To benchmark ProTEUS against MULTI-evolve, we compared their zero-shot rankings across 13 held-out single-substitution DMS assays (fig. S2B) and evaluated variant selection under matched selection budgets. Among the 10 highest-ranked candidates, 75.4% selected by ProTEUS and 64.6% selected by MULTI-evolve had fitness at or above the assay mean (Fig. 2F), showing ProTEUS's superiority compared to ensemble methods. These results support the use of the pretrained world model to prioritize candidate variants on new protein backgrounds, providing a starting point for experimental evolution before target-specific feedback becomes available.

To assess how well the model can leverage experimental measurements on a particular activity to further guide engineering, we evaluated supervised fine-tuning on the held-out DMS assays using within-assay fivefold cross-validation. On held-out variants, ProTEUS achieved a mean Spearman correlation of 0.549, compared with 0.367 for ProteinNPT (*34*), the next-ranked method (Fig. 2G). Lastly, to evaluate whether the model can efficiently use low-N measurements to guide engineering in an active learning fashion, we simulated directed evolution on 10 single-mutant DMS datasets. ProTEUS and EVOLVEpro each selected 30 variants per round from the same candidate library over five rounds. Measurements for selected variants were revealed and used to update the model before the next round. By round 5, ProTEUS reached a mean normalized best fitness of 0.741, compared with 0.555 for EVOLVEpro (Fig. 2H). Overall, these computational benchmarks support ProTEUS as a state-of-the-art model for zero-shot recommendation of mutations and supervised fine-tuning on target specific activity data.

**Efficient physics-constrained model supports multi-mutant screening**

To efficiently sample a large number of mutations in silico, especially for higher-order variants that differ in multiple residues, ProTEUS would need to generate high quality structures with atomic level precision and minimal computational cost per structure to supply as input (*32*). Current frontier structural prediction models like AlphaFold3 still suffer from poor physical validity and a relatively large computational budget due to the use of diffusion modules (*42*), so we developed a fast all-atom predictor that combined two denoising steps with explicit geometric constraint enforcement (*32*).

The predictor retained the atom-attention encoder, MSA module and PairFormer of Boltz-1 (*43*), which provided conditioning features for a diffusion network initialized from its released weights. The network generated provisional all-atom coordinates, and a differentiable Gauss–Seidel projection corrected geometric and stereochemical violations (Fig. 3A-B). At inference, projection followed the second denoising update. The learned predictor thus recovered molecular conformation, while the projection enforced local constraints on the resulting coordinates. The projection balanced correction of constraint violations against displacement from the predicted structure. Seven constraint families covered nonbonded inter-chain separation, tetrahedral chirality, E/Z double-bond stereochemistry, carbon double-bond planarity, ligand internal distances, symmetric-chain overlap and inter-chain covalent bonds (Fig. 3B; table S3). The solver exploited constraint locality by updating affected coordinates in place and using each correction in subsequent updates. Twenty Gauss–Seidel sweeps resolved coupled constraints without assembling a large global system (*32*). Constraints without shared atoms were processed in parallel on the GPU to reduce the cost of geometric correction.

We included projection during training so that the predictor learned to generate coordinates compatible with the subsequent correction. The Boltz-1-initialized network was fine-tuned on approximately 180,000 Protein Data Bank structures, with the denoising loss evaluated on projected coordinates. Implicit differentiation propagated gradients through the projection without storing the full sequence of solver iterations. In the reported CASP15 ablation (*32*), fine-tuning through projection increased complex LDDT from 0.5889 to 0.6321 relative to inference-only projection. Both settings passed all seven validity checks. The fine-tuned two-step predictor approached the complex LDDT of 0.6406 obtained with the 200-step Boltz-1 reference. Training through projection therefore helped retain structural accuracy under a reduced denoising budget.

We evaluated structural accuracy and geometric validity across CASP15, Test, PoseBusters, AF3-AB, dsDNA and RNA–protein benchmarks. Global accuracy, measured by complex LDDT (*44*) and TM-score (*45*), was broadly comparable to the evaluated baselines, and all projected structures passed the seven validity checks (Fig. 3C), when baseline predictions did not consistently satisfy these checks. The complete inference procedure, including denoising and projection, achieved a median 9.4-fold speedup over 200-step Boltz-1 on CASP15 (*32*) (Fig. 3D). Across five atom-count bins, AlphaFold 3 mean runtimes were 7.81–10.81 times higher than those of our predictor. In the representative protein–ligand complex PDB 7WCF, AlphaFold 3 predicted a ligand–side-chain separation of 1.41 Å, whereas our predictor yielded 4.44 Å, reflecting more accurate resolution of local geometry that is critical for a variety of downstream tasks including molecular dynamics and protein engineering (Fig. 3E).

## Higher-order mutation activity prediction

Virtual exploration of higher-order variants requires predicting functional outcomes beyond the mutation order covered by existing measurements. We therefore tested whether ProTEUS could generalize from measured lower-order variants to higher-order combinations. In a K-vs-rest benchmark across 15 multi-mutant DMS datasets, models were trained on variants carrying up to K substitutions and tested on variants carrying more than K (*46*). For this task, we compared ProTEUS directly with MULTI-evolve for higher-order fitness prediction (Fig. 3F). Across 40 datasets-K comparisons, ProTEUS achieved a mean Spearman correlation of 0.448, compared with 0.295 for MULTI-evolve. Among these comparisons, ProTEUS achieved higher correlation in 80%, reflecting a more information-rich joint embedding space ProTEUS uses.

We then evaluated whether predictions updated with new measurements could guide successive rounds of higher-order variant selection. We compared ProTEUS with EVOLVEpro and MULTI-evolve in simulations of progressive directed evolution on 17 DMS datasets. We also included an additive model fitted to accumulated measurements, providing a baseline for variant selection based on individual substitution effects without interaction terms (*47*). Each method selected 30 variants per round from the same candidate library over five model-guided rounds following an initial batch. Measurements for selected variants were revealed and used to update the model before the next round. ProTEUS achieved the highest mean normalized best fitness in each round (Fig. 3G), extending its predictive performance to iterative exploration of mutation combinations.

Based on these results, we finalized ProTEUS's closed-loop evolution strategy by prompting ProTEUS to zero-shot rank an initial single-mutant batch. Subsequent nominations combined zero-shot scores, updated fitness predictions and fitness ranks derived from all available experimental measurements (fig. S4; table S4). The highest-ranked eligible, untested variants were selected, with the maximum allowed mutation order retained or increased by one. Measurements from each batch updated the model and the experimental evidence available for subsequent nominations.

## Agentic Closed-loop evolution of levansucrase with ProTEUS

To smoothly integrate ProTEUS with wet-lab and the growing capabilities of LLMs, we built an agentic workflow to automate the end to end closed-loop protein evolution (Fig. 4A). An agent coordinates sequence candidates generation, fitness prediction, mutation nomination, and incorporates experimental feedback to update the model for subsequent rounds. Beyond workflow automation, the agent supports scientific assessment of the mutations proposed by ProTEUS. ProTEUS prioritizes the top-k candidate mutation sites, and the agent combines ProTEUS scores with LLM-based analysis of the physicochemical implications of the proposed substitutions to generate a rationale and confidence assessment for each nominated variant. These assessments accompany the ranked candidates in a human-in-the-loop review step, helping scientists evaluate the proposed substitutions and make informed choices about mutation sites for experimental testing. As modern LLMs get increasingly knowledgeable about diverse proteins, these assessments could also be incorporated as part of scoring function in the future (*48*).

To evaluate ProTEUS-guided evolution, we followed the predefined variant-nomination strategy from our computational benchmarks. For the two closed loop engineering campaigns shown here, agent-generated rationales and confidence assessments served as interpretive annotations, without human adjustment of candidate rankings or batch selection based on these assessments. We first deployed agentic ProTEUS to engineer levansucrase (SacB) from *Bacillus subtilis*. Levansucrase transfers fructosyl groups from sucrose to synthesize levan, releasing glucose (*49*). Levan's use as a viscosifier, stabilizer and gelling agent in a variety of food applications motivates efforts to improve its enzymatic activity and stability (*50*). To generate high throughput levansucrase activity and expression data, we cloned mutations suggested by ProTEUS into a his tagged vector for expression in *E. coli* and then measured glucose production from sucrose breakdown to infer catalytic activity of the enzyme (Fig. 4A). In addition, we can measure the expression of the enzyme directly by an anti-His tag ELISA.

Using this setup, we performed a multi-objective engineering by considering the product of expression and catalytic activity as the reward function since mutations that only improve activity but compromise expressions and stability are not useful in downstream applications. ProTEUS started with an initial zero-shot recommendation of 30 single mutants, and laboratory testing revealed that Q34N mutant within this batch had a 4.5-fold higher activity than the WT protein, reaffirming the ability to zero-shot identify high fitness candidates as seen in previous computational benchmarks (Fig. 4B-C). Subsequently, ProTEUS used this data to suggest higher-order mutants in a round wise fashion (Fig. 4B). The next round identified Q34N:G240W and Q34N:S359F with activities of 8.94-fold and 8.91-fold WT, respectively (Fig. 4C). Both double-mutant parents were carried forward by ProTEUS, allowing proper exploration of two backgrounds with similar measured activities. In the third experimental batch, ProTEUS suggested 18 triple mutants derived from Q34N:G240W and 12 derived from Q34N:S359F (table S5). In the third round, G23A:Q34N:G240W reached 14.03-fold WT activity, 1.57 times that of its parent, while Q34N:M328L:S359F reached 12.38-fold WT activity (Fig. 4C, fig. S5A; table S10). Q34N:G240W:Y436E also exceeded the preceding round's best variant, reaching 10.56-fold WT activity. Similar to our computational benchmarks, ProTEUS can properly identify synergistic mutations during exploration of higher-order mutations. Overall, ProTEUS was able to find variants with >10 fold higher activity than the starting WT with a budget of <100 experimental tests, highlighting its ability to efficiently leverage experimental data for closed-loop evolution.

To better understand the mutations explored in this closed-loop evolution, we examined how the scores used to evaluate candidates relate to measured activity within each experimental batch. Initial zero-shot scores had a Spearman correlation of −0.24 with measured activity. It is worth noting that even though the initial zero-shot batch of 30 mutants properly identified high fitness variants like Q34N, the overall low global spearman correlation still highlights the divergence between model inferred protein fitness and the specific enzymatic activity and expression being optimized in this campaign. Composite scores that incorporated experimental data in the subsequent rounds were positively correlated with activity, with correlations of 0.24 in round 1 and 0.38 in round 2, reflecting the model’s ability to leverage experimental data to align its output to the specific engineering objective (Fig. 4D). Structural mapping placed the selected substitutions in several regions of the protein (Fig. 4E). In the WT structure, Q34 and G240 were 29.5 Å and 10.5 Å from the superposed sucrose, respectively. The alternative branch included S359, 7.6 Å from sucrose, and M328, 19.0 Å away. Similar to previously reported, these mutations do not fall in the key catalytic active site of the enzyme and thus likely contributed to the overall improvement in activity through other complementary mechanisms (*3*).

To examine the candidate states explored during evolution, we mapped WT and the 90 assayed variants onto a shared t-SNE projection containing 366,836 library representations (*51*). The overlaid lineage traced successive additions of Q34N, G240W and G23A in this visualization of the model latent space (Fig. 4F). We then assessed how the original 768-dimensional representation distinguished candidates within each final-round parental background. Among pairs differing at two residues, cosine distances ranged from 0.00151 to 0.03240 on Q34N:G240W and from 0.00310 to 0.03195 on Q34N:S359F (table S9). Within the Q34N:G240W background, 6 principal components were enough to capture 90% of the model representation variance, compared with 14 for the amino-acid one-hot representations used in MULTI-evolve (table S9).

We next assessed the novelty of the mutations suggested by ProTEUS. In an alignment of 2,451 homologous sequences, serine was the consensus residue at position 34, occurring in 63 of 188 aligned sequences (33.5%). However, Q34S had only 0.56-fold WT activity, whereas Q34N reached 4.50-fold WT activity despite asparagine occurring in only 7 of 188 sequences (3.72%). This highlights ProTEUS's ability to deduce novel mutations that are potentially beneficial to protein activity, without overrelying on evolutionary information. In the closed-loop evolution campaigns, all three substitutions in the final lead introduced non-consensus residues. The residues introduced by G23A and G240W occurred in 6 of 184 (3.26%) and 4 of 1,258 (0.32%) aligned sequences, respectively. The originality ratios for G23A, Q34N and G240W were 0.04, 0.11 and 0.01 (Fig. 4G). Across the campaign, the 89 distinct substitutions at 65 positions had a median introduced-residue frequency of 1.19%, and 22 were unobserved at their respective alignment positions. These measurements showed that the search retained beneficial substitutions that were uncommon among homologs while pursuing further improvements on experimentally supported backgrounds.

### Closed-loop evolution identifies improved ALDH

To better understand whether ProTEUS's capability generalizes beyond one enzyme tested, we next applied ProTEUS to engineer aldehyde dehydrogenase (ALDH) from *Cupriavidus pinatubonensis*. ALDH catalyzes the oxidation of aldehydes to carboxylic acids. Human genetics have shown that mutation in the ALDH gene can lead to build up of acetaldehyde and is implicated with increased risk of cardiovascular disease and cancer (*52–54*). To identify a more potent ALDH enzyme, we established a high-throughput assay to measure total ALDH activity in E. coli lysates by detecting NADH generated during acetaldehyde oxidation with a coupled reaction that converted a colorless probe into a product absorbing at 450 nm.

We prompted ProTEUS to nominate thirty variants per round for five rounds, totaling 150 distinct variants (table S11). The search began with zero-shot nomination of single substitutions and progressed to double, triple, and quadruple mutants (Fig. 5A). Activity measurements were returned to ProTEUS between rounds to update the fitness predictor and inform subsequent nominations. During this evolution campaign, the initial zero-shot prediction round identified L358I, with 2.02 times higher than WT activity (Fig. 5B). Building on L358I, subsequent rounds leveraging experimental datapoints identified A116K and H332N, with the triple mutant reaching 6.08 times WT activity (Fig. 5A-B, fig. S5B). In the final round, we identified S103V:A116K:H332N:L358I with 6.85-fold WT activity (Fig. 5B; table S5). Because the high-throughput screen measured total activity without normalizing protein abundance, we used an anti-His tag ELISA kit to quantify the enzyme concentration per variant, and calculated the normalized enzyme catalytic rate fold change. This showed that the best mutant in the final round (round 4) had a 7.90 fold improvement of the catalytic rate over WT with other variants showing different combinations of enzyme abundance and concentration-normalized activity (fig. S5C). Lastly, we purified the WT, L358I and the final quadruple mutant and measured reaction time courses using equal amounts of purified protein. The final variant had a higher absorbance slope than WT and L358I (Fig. 5C), supporting increased activity with equal amounts of purified protein. To better understand ProTEUS's engineering trajectory, we examined the relationship between scores and measured activity within each experimental batch. Similar to the levansucrase campaign, zero-shot scores showed little association with activity in the initial batch, whereas composite scores were positively correlated with activity in subsequent rounds. The composite score reached a Spearman correlation of 0.66 in the final round among the 30 tested variants (Fig. 5D). We then mapped the top mutations onto the structure of the enzyme and again found that all four altered positions were distal to the annotated catalytic site, lying more than 13 Å from the annotated catalytic residues (Fig. 5E).

To examine how the learned state space represented the experimental trajectory, we analyzed the organization and measured activities of ALDH variants in the ProTEUS embedding (tables S6-7). We projected WT and the 150 assayed variants into a shared t-SNE map containing 354,144 candidate representations (Fig. 5F). Initial single-mutant selections occupied multiple regions, whereas subsequent higher-order variants localized around selected parental backgrounds. Lineage mapping traced successive additions of L358I, A116K and H332N. The final two batches explored alternative fourth substitutions on A116K:H332N:L358I, with S103V yielding the final lead. We then tested whether neighboring representations were associated with similar activities among the 60 four-substitution variants sharing the A116K:H332N:L358I background. Each variant was compared with its five nearest neighbors from the same batch in the original 768-dimensional space. Mean absolute activity differences were 1.69 between embedding neighbors and 1.88 for sequence-distance-matched comparisons, with WT activity set to 1. This represented a 10.1% reduction (two-sided permutation $P = 0.046$; table S6). Three- and ten-neighbor analyses showed smaller reductions of 6.7% and 6.0%, respectively, with statistical support varying across neighborhood sizes (table S6). Within the A116K:H332N:L358I background, neighboring variants in the ProTEUS embedding showed smaller activity differences than sequence distance matched comparisons.

Lastly, to assess the evolutionary prevalence of the substitutions explored during ALDH evolution, we analyzed an alignment of 19,704 homologous sequences (table S8). Across 148 distinct substitutions at 104 positions, the median substituted-residue frequency was 0.59%, and 10 substitutions were unobserved at their respective alignment positions (Fig. 5G). In the final lead, two mutations S103V and A116K were rare and novel mutations whereas the original L358I introduced in the zero-shot discovery and H332N were frequently observed residues during evolution. This again highlights ProTEUS's ability to integrate multi-modal information both from evolution and experimental data to pick up both common and rare beneficial mutations.

# Discussion

Efficiently finding a set of mutations that will push the protein toward a desired phenotype is one of the holy grails of molecular engineering. In the vast space of $20^L$ possible mutations that is not possible to experimentally characterize, any computational methods that can identify a path that leads to desired activity is of great value, especially for activity not explored in nature but critical for biotechnology applications. In the age of generative protein design, many models, especially large pretrained foundation models, have been developed to rank protein variants in a zero-shot fashion. However, this class of method in general suffers from the fact that many protein activities are not well represented in evolution. In contrast, active learning methods like EVOLVEpro demonstrated that pretrained protein representations can guide protein engineering (single mutant) by using relatively little experimental data points to align the model directly to experimental activity of interest (*3*), and MULTI-evolve has further shown that it is possible to guide engineering of higher-order variants (*1*). These two models highlight the utility of active learning to align model output with wet lab-measured protein activity, but still suffer from the lack of a good unified representation of protein variants to efficiently leverage experimental data and guide engineering toward higher order mutations with complex epistasis (*47*).

With an ever increasing capability of DNA synthesis, sequencing and more powerful deep learning models, we propose a new way to look at this task, a world model that connects the virtual evaluation of proposed mutations with experimental learning. In essence, such a world model would need a high quality latent space to understand and reason a given mutation's impact on a protein phenotype and leverage experimental measurement in an efficient manner to guide a model's search inside this vast sequence space. To build a prototype of this world model, we leveraged a MutationFormer that integrates sequence, evolutionary and structural context to learn a joint embedding space capable of understanding sparse mutations' impact on protein activity, with a physics-constrained structural predictor that significantly accelerates generation of candidate-specific structural inputs. ProTEUS achieves higher accuracy on both zero-shot ranking of protein variants and supervised fine tuning tasks where we test whether the model can efficiently learn from experimental data to extrapolate a target-specific activity landscape. Our experimental validation on ALDH and levansucrase presents preliminary evidence for the utility of such a model to guide protein engineering in a closed-loop fashion.

There are numerous ways ProTEUS could be further improved. First, the pretraining of ProTEUS leveraged ProteinGym, which so far hosts around 2.5 million mutational measurements (*29*). In many deep learning models, especially modern day LLMs, more data during pretraining generally leads to better model performance and generalization. This scaling law both in terms of pretraining data and model size has been the driver of breakthroughs in areas like natural language processing, computer vision, and is also seen in protein structural prediction and protein language model (*55–59*). The data-scaling analysis showed that ProTEUS predicted held-out variant fitness more accurately as the amount of functional pretraining data increased (Fig. 2C). Protein mutation datasets typically are very noisy with strong variance even for the same protein from different sources due to the sensitive nature of different assays and measurement conditions, and thus have far fewer high quality curation compared to other domains. However, with initiatives like the ProteinGym (*29*) and better standardization of high throughput protein activity datasets similar to early efforts in the protein structural domains that resulted in protein data bank, we should expect an accumulation of high quality mutation datasets on diverse protein activities in the coming decade, and correspondingly greater model performance and generalization for pretrained models like ProTEUS. With the rise of LLMs for text processing, it is also plausible that a comprehensive mining of literatures for scattered mutation experiments will lead to collection of activity dataset for pretraining. Together, these directions represent exciting new avenues to further test the scaling behavior of protein engineering world models.

Second, we have also explored the axis of active learning size in this study (fig. S3). Similar to previous reports, larger amounts of experimental data in a closed-loop evolution campaign lead to better performance. The agentic feature of ProTEUS and the rise of wet lab automation thus represent a new exciting frontier where autonomous collections of wet lab experimental datapoints are guided by the model to continuously update its weight for optimal search of the protein sequence landscape (*60–62*). These data points not only guide a model's current engineering campaign on a specific target phenotype, but also could be used as training data in future pre-training, creating a positive feedback loop.

ProTEUS puts the protein engineering task into a mutation-resolved world-model framework for experimental evolution, complementing efforts to learn general representations of protein biology (*56*). In this framework, an informative joint embedding space allows the model to reason about a mutation's impact on a desired protein phenotype through understanding laboratory experiments. This provides a foundation for protein engineering in which virtual exploration and experimental learning advance together toward a desired function, broadly enabling the synthetic biology and molecular engineering community to select for protein sequences with the desired phenotype efficiently.

# Acknowledgements

We would like to thank members of the Jiang Lab, Le Lab, Chen Lab, Matusik lab and the Jin Lab for helpful discussion on the manuscript.

# Funding

This research was funded by Foundry Biosciences Inc.

# Author Contributions

Conceptualization: K.J.; Methodology: Y.Z., M.G. K.J..; Investigation: Y.Z., M.G., Y.C., G.X.; Visualization: Y.Z., M.G.; Funding acquisition: K.J.; Project administration: K.J., J.C.; Supervision: K.J.; Writing: all authors.

# Competing Interests

K.J. and J.C. are co-founder of Foundry BioSciences Inc. Y.Z. was an employee of Foundry BioSciences Inc. All other authors declare no conflicts of interest. Patents have been filed related to aspects of this work.

# Data Availability

The deep mutational scanning datasets used for pretraining and evaluation were obtained from ProteinGym (*29*), CombinGym (*46*), MaveDB (*33*) and the published sources described in Methods. Structural data are available from the RCSB Protein Data Bank (*63*). Sequences and computational details and analyses are provided in the Supplementary Materials. Models and code are available at https://github.com/yikunpku/ProTEUS.

# List of Supplementary Materials:

**Materials and Methods**
**Supplementary Text**
**Figs. S1 to S5**
**Tables S1 to S11**

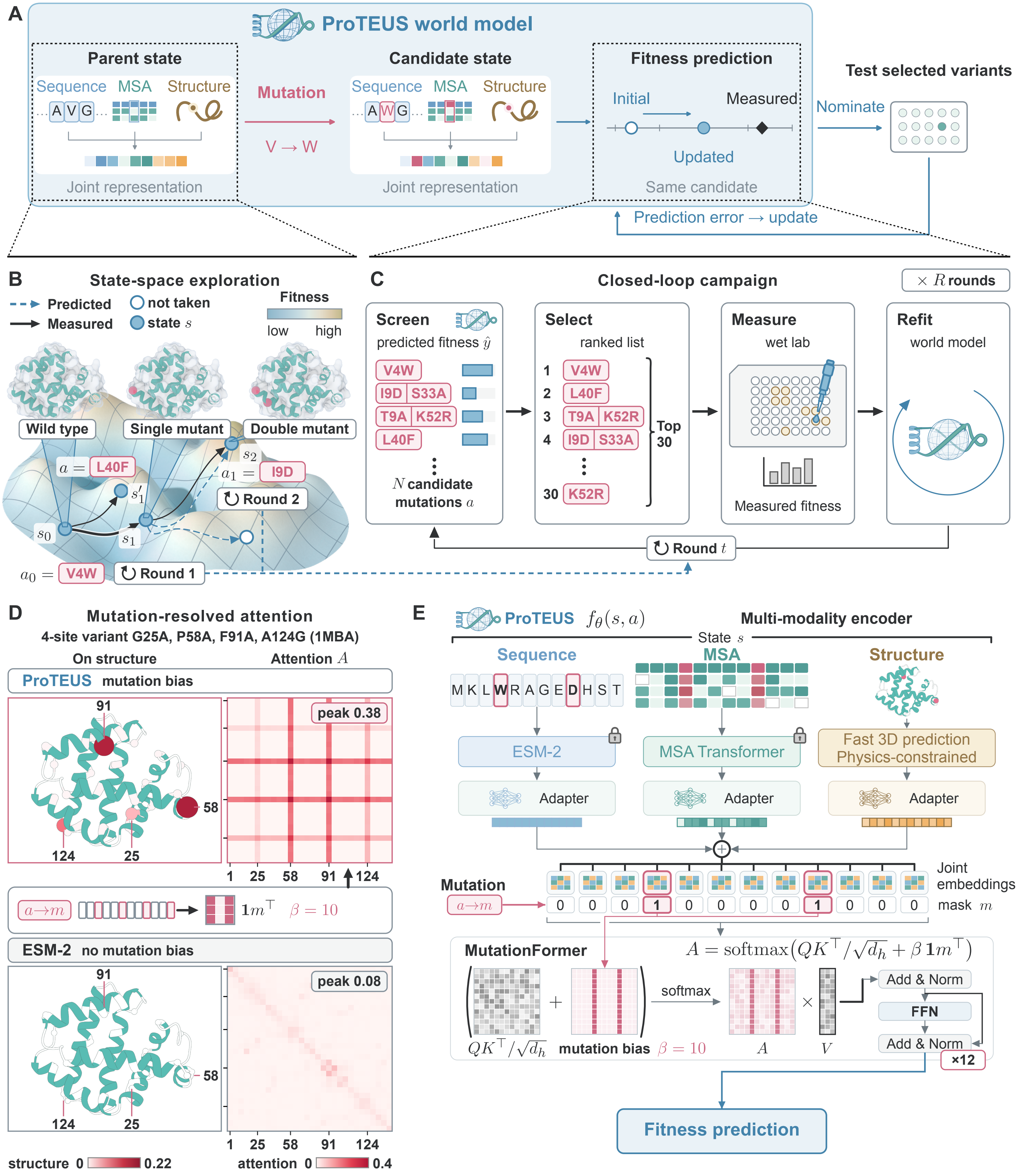
A
ProTEUS world model
Parent state
Sequence MSA Structure
…AVG…
Joint representation
Mutation
V → W
Candidate state
Sequence MSA Structure
…AWG…
Joint representation
Fitness prediction
Initial
Measured
Updated
Same candidate
Test selected variants
Nominate
Prediction error → update
B
State-space exploration
Predicted
Measured
not taken
state $s$
Fitness
low high
Wild type
Single mutant
Double mutant
$a$ = L40F
$s_2$
$a_1$ = I9D
$s_1'$
Round 2
$s_0$
$s_1$
$a_0$ = V4W
Round 1
C
Closed-loop campaign
× $R$ rounds
Screen
predicted fitness $\hat{y}$
V4W
I9D S33A
T9A K52R
L40F
$N$ candidate mutations $a$
Select
ranked list
1 V4W
2 L40F
3 T9A K52R
4 I9D S33A
30 K52R
Top 30
Measure
wet lab
Measured fitness
Refit
world model
Round $t$
D
Mutation-resolved attention
4-site variant G25A, P58A, F91A, A124G (1MBA)
On structure
Attention $A$
ProTEUS mutation bias
91
58
124
25
peak 0.38
1 25 58 91 124
$a \to m$
$\mathbf{1}m^\top$ $\beta = 10$
ESM-2 no mutation bias
peak 0.08
structure 0 0.22
attention 0 0.4
E
ProTEUS $f_\theta(s, a)$
Multi-modality encoder
State $s$
Sequence
MSA
Structure
M K L W R A G E D H S T
ESM-2
MSA Transformer
Fast 3D prediction Physics-constrained
Adapter
Adapter
Adapter
Mutation
$a \to m$
Joint embeddings
mask $m$
0 0 0 1 0 0 0 0 1 0 0 0
MutationFormer
$A = \mathrm{softmax}(QK^\top/\sqrt{d_h} + \beta\,\mathbf{1}m^\top)$
$QK^\top/\sqrt{d_h}$
mutation bias $\beta = 10$
softmax
$A$
$V$
Add & Norm
FFN
Add & Norm
×12
Fitness prediction

Figure 1

**Figure 1. A mutation-resolved world model for closed-loop protein evolution**

(A) Overview of the ProTEUS world model. Parental and candidate states combine sequence, evolutionary and structural information in a joint latent space. Mutations specify sequence changes, predicted fitness guides nomination, and measurements support subsequent prediction updates. The sequence segments and fitness estimates are illustrative. (B) State-space exploration: Protein variants are represented as states, and substitutions define actions between parental and candidate sequences. Dashed and solid arrows distinguish predicted and experimentally evaluated outcomes on a schematic fitness landscape. (C) Closed-loop campaign connecting virtual screening, variant nomination, laboratory measurement and model updating over successive rounds. (D) Illustration of mutation-biased attention in an Aplysia limacina myoglobin variant carrying G25A, P58A, F91A and A124G (PDB: 1MBA) *(64)*. The schematic illustrates how an explicit mutation bias emphasizes substituted positions in ProTEUS. Structural projections and attention maps show the same 30 residues, including the four substituted positions. (E) Multimodal architecture. Sequence, MSA and structural features are aligned into joint residue embeddings. MutationFormer integrates these features with attention biased toward substituted positions, producing mutation-resolved embeddings for fitness prediction.

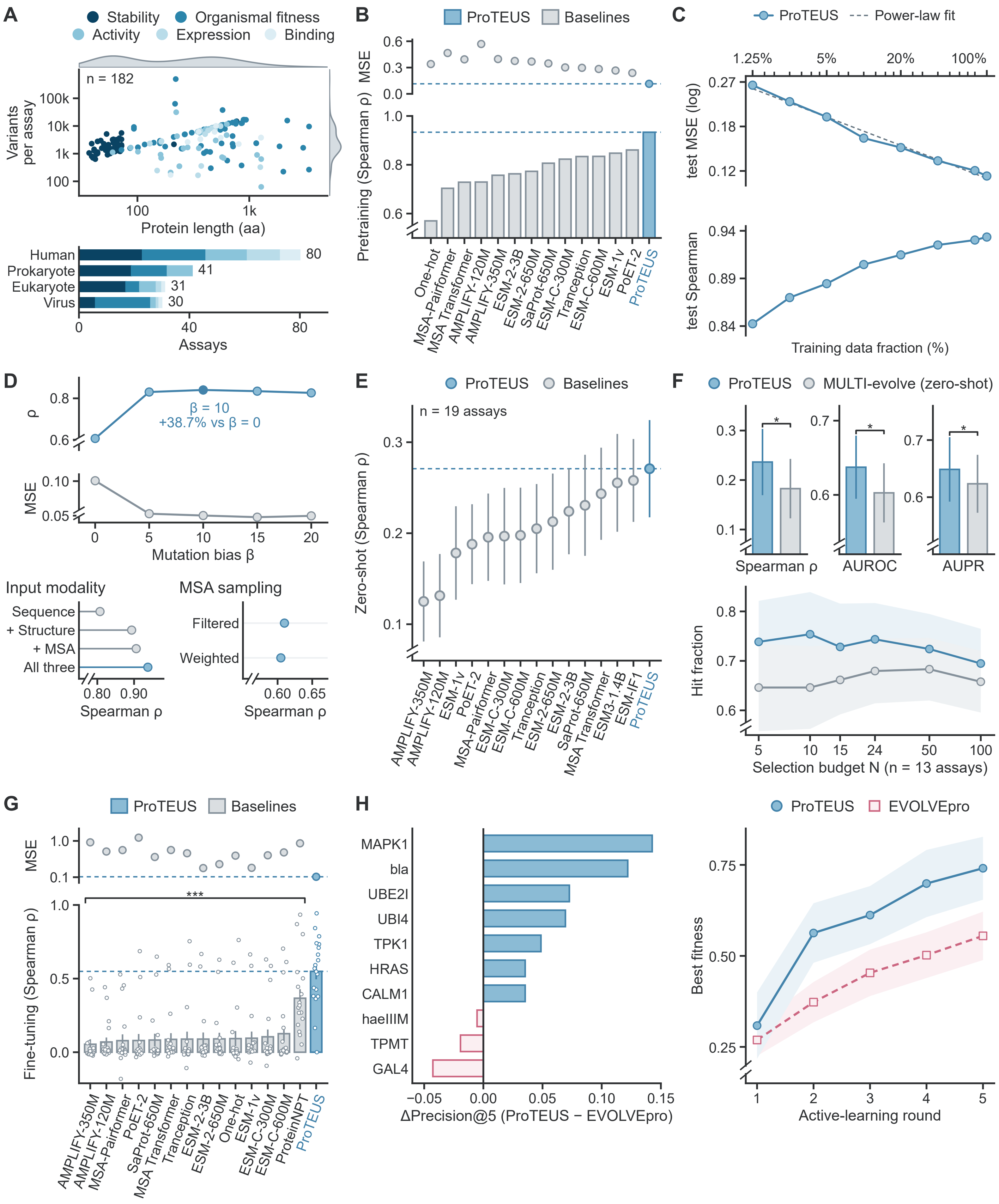
A
Stability
Organismal fitness
Activity
Expression
Binding
n = 182
Variants per assay
100
1k
10k
100k
Protein length (aa)
Human
Prokaryote
Eukaryote
Virus
80
41
31
30
0
40
80
Assays
B
ProTEUS
Baselines
MSE
0.6
0.3
0.0
Pretraining (Spearman ρ)
1.0
0.8
0.6
One-hot
MSA-Pairformer
MSA Transformer
AMPLIFY-120M
AMPLIFY-350M
ESM-2-3B
ESM-2-650M
SaProt-650M
ESM-C-300M
Tranception
ESM-C-600M
ESM-1v
PoET-2
ProTEUS
C
ProTEUS
Power-law fit
1.25%
5%
20%
100%
test MSE (log)
0.27
0.18
0.12
test Spearman
0.94
0.89
0.84
Training data fraction (%)
D
ρ
0.8
0.6
β = 10
+38.7% vs β = 0
MSE
0.10
0.05
0
5
10
15
20
Mutation bias β
Input modality
Sequence
+ Structure
+ MSA
All three
0.80
0.90
Spearman ρ
MSA sampling
Filtered
Weighted
0.60
0.65
Spearman ρ
E
ProTEUS
Baselines
n = 19 assays
Zero-shot (Spearman ρ)
0.3
0.2
0.1
AMPLIFY-350M
AMPLIFY-120M
ESM-1v
PoET-2
MSA-Pairformer
ESM-C-300M
ESM-C-600M
Tranception
ESM-2-650M
ESM-2-3B
SaProt-650M
MSA Transformer
ESM3-1.4B
ESM-IF1
ProTEUS
F
ProTEUS
MULTI-evolve (zero-shot)
0.3
0.2
0.1
Spearman ρ
0.7
0.6
AUROC
0.7
0.6
AUPR
Hit fraction
0.8
0.7
0.6
5
10
15
24
50
100
Selection budget N (n = 13 assays)
G
ProTEUS
Baselines
MSE
1.0
0.1
***
Fine-tuning (Spearman ρ)
1.0
0.5
0.0
AMPLIFY-350M
AMPLIFY-120M
MSA-Pairformer
PoET-2
SaProt-650M
MSA Transformer
Tranception
ESM-2-3B
ESM-2-650M
One-hot
ESM-1v
ESM-C-300M
ESM-C-600M
ProteinNPT
ProTEUS
H
MAPK1
bla
UBE2I
UBI4
TPK1
HRAS
CALM1
haeIIIM
TPMT
GAL4
−0.05
0.00
0.05
0.10
0.15
ΔPrecision@5 (ProTEUS − EVOLVEpro)
ProTEUS
EVOLVEpro
Best fitness
0.75
0.50
0.25
1
2
3
4
5
Active-learning round

# Figure 2

**Figure 2. Mutation-resolved representations support fitness prediction and variant selection**

(A) Distribution of the 182 deep mutational scanning (DMS) assays used for pretraining by protein length, number of variants, taxonomic group, and measured phenotype. Each point in the scatter plot represents an assay. (B) Fitness prediction for variants held out during pretraining. Bars show Spearman correlation and open circles show mean squared error (MSE). (C) Data scaling of functional pretraining. Test MSE (upper) and Spearman correlation (lower) are shown for eight training-data fractions from 1.25% to 100%. Both plots share a logarithmic horizontal axis; the MSE vertical axis is logarithmic and the Spearman vertical axis is linear. Blue points show measurements, and solid blue lines connect observations. The grey dashed line shows the power-law fit to MSE (exponent, −0.187; $R^2$ in log space, 0.989). (D) Effects of mutation-attention bias, input modalities, and MSA sampling on fitness prediction for the same dataset in (B). The mutation-bias experiment reports Spearman correlation and MSE; modality and sampling comparisons report Spearman correlation. (E) Zero-shot fitness prediction across 19 held-out assays. Points and error bars indicate mean ± standard error of the mean (SEM) across assays. (F) First-round nomination of single mutants by ProTEUS and MULTI-evolve (ensemble model prediction) rank fusion across 13 held-out DMS assays. Upper panels show Spearman correlation, area under the receiver operating characteristic curve (AUROC), and area under the precision – recall curve (AUPR). Lower panel shows the fraction of selected variants with fitness at or above the assay mean, with at most one substitution selected per residue position. Bars or lines indicate means; error bars and shading indicate SEM. P values are from one-sided paired Wilcoxon signed-rank tests across assays; *, $p < 0.05$. (G) Supervised fine-tuning using fivefold cross-validation. Spearman correlations are shown as means ± SEM across assays, with individual assay values overlaid; open circles show mean MSE. One-sided paired t-tests were used for the displayed comparisons; ***, $p < 0.001$. (H) Five rounds of simulated directed evolution on 10 single-mutant DMS datasets, with 30 variants selected per round by each method. Left, dataset-specific differences in Precision@5 (ProTEUS minus EVOLVEpro), defined as the fraction of selected variants whose measured fitness falls within the top 5% of the dataset. Positive values indicate higher Precision@5 for ProTEUS. Right, the highest fitness recovered up to each round, normalized within each dataset. Lines and shaded bands indicate the mean and SEM across the 10 datasets, respectively.

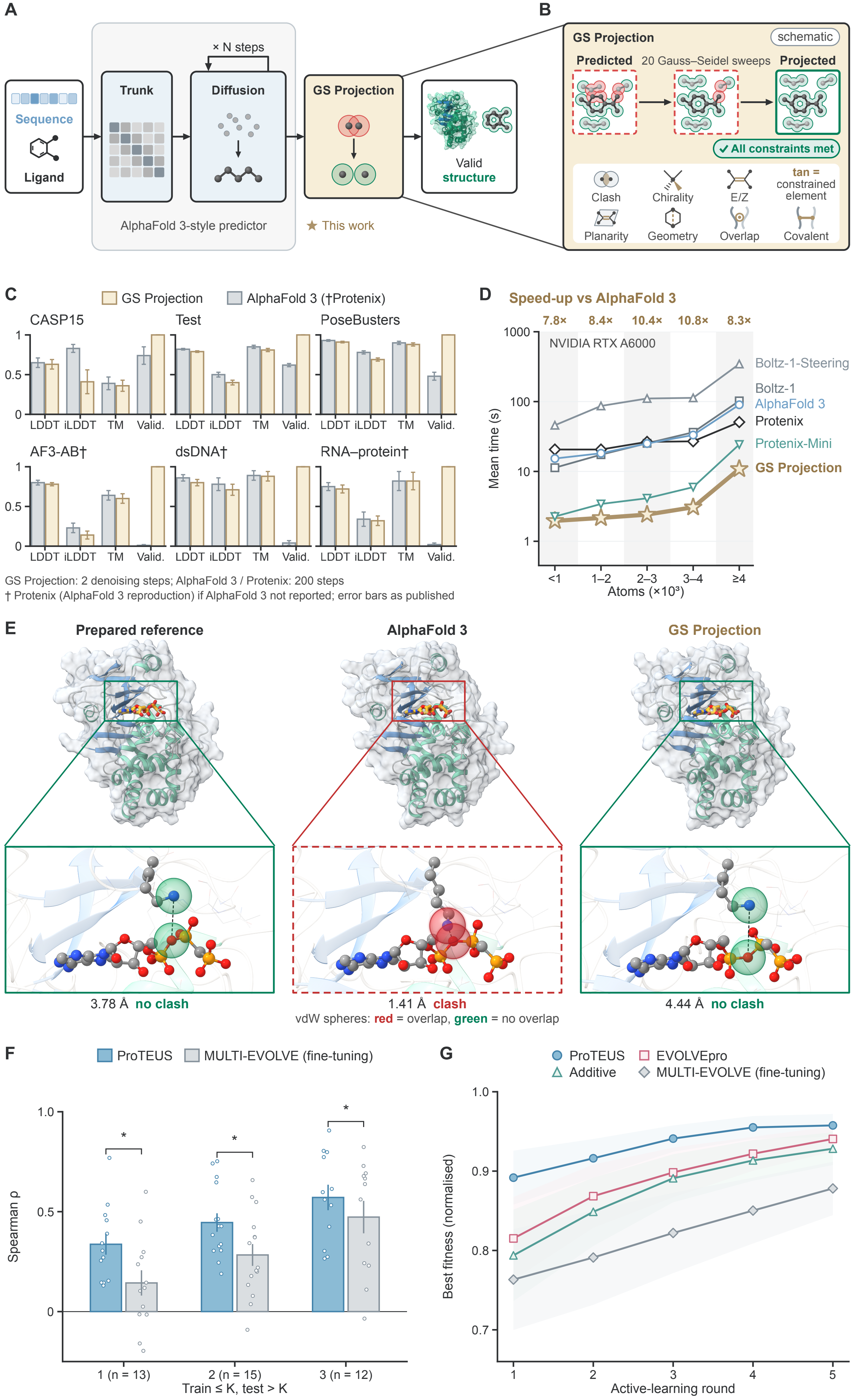

A
× N steps
Sequence
Ligand
Trunk
Diffusion
AlphaFold 3-style predictor
GS Projection
★ This work
Valid structure
B
GS Projection
schematic
Predicted
20 Gauss–Seidel sweeps
Projected
✓ All constraints met
Clash
Chirality
E/Z
tan = constrained element
Planarity
Geometry
Overlap
Covalent
C
GS Projection
AlphaFold 3 (†Protenix)
CASP15
Test
PoseBusters
AF3-AB†
dsDNA†
RNA–protein†
LDDT iLDDT TM Valid.
GS Projection: 2 denoising steps; AlphaFold 3 / Protenix: 200 steps
† Protenix (AlphaFold 3 reproduction) if AlphaFold 3 not reported; error bars as published
D
Speed-up vs AlphaFold 3
7.8× 8.4× 10.4× 10.8× 8.3×
NVIDIA RTX A6000
Mean time (s)
Boltz-1-Steering
Boltz-1
AlphaFold 3
Protenix
Protenix-Mini
GS Projection
<1 1–2 2–3 3–4 ≥4
Atoms (×10³)
E
Prepared reference
AlphaFold 3
GS Projection
3.78 Å no clash
1.41 Å clash
4.44 Å no clash
vdW spheres: red = overlap, green = no overlap
F
ProTEUS
MULTI-EVOLVE (fine-tuning)
Spearman ρ
1 (n = 13)
2 (n = 15)
3 (n = 12)
Train ≤ K, test > K
G
ProTEUS
EVOLVEpro
Additive
MULTI-EVOLVE (fine-tuning)
Best fitness (normalised)
Active-learning round

# Figure 3

**Figure 3. Physics-constrained structure prediction supports exploration of higher-order variants**

(A) A schematic of the overall structural prediction module, where all-atom prediction is followed by geometric projection. Sequence and ligand inputs provide conditioning features for diffusion-based coordinate generation; Gauss–Seidel projection corrects geometric and stereochemical violations. (B) Projection through 20 Gauss–Seidel sweeps. Seven constraint families address steric clashes, tetrahedral chirality, E/Z stereochemistry, carbon double-bond planarity, ligand internal distances, symmetric-chain overlap and inter-chain covalent bonds. (C) Structural accuracy and physical validity across six biomolecular benchmarks (32). The two-step predictor is compared with 200-step AlphaFold 3 (42) on CASP15, Test and PoseBusters (65), and 200-step Protenix (66) on AF3-AB, dsDNA and RNA–protein (67). Metrics are complex LDDT, interface LDDT, TM-score and the fraction of predictions satisfying all seven validity checks. Scores were first averaged across generated samples for each protein. Bars show the mean across proteins, and error bars indicate 95% confidence intervals. (D) Mean wall-clock runtime on CASP15 across five atom-count bins. Comparators include AlphaFold 3, Boltz-1, Boltz-1-Steering, Protenix and Protenix-Mini. The headline reports speedup relative to Boltz-1; the shaded band and labels below the stars show speedup relative to Protenix-Mini. Speedup is the ratio of mean runtimes within each bin, and the median is calculated across the five bin-specific ratios. (E) Representative ligand–protein contact in PDB 7WCF (68). Matched views show the ACP O3A–Lys Nζ distance in the prepared reference, AlphaFold 3 prediction and physics-constrained prediction. Van der Waals spheres indicate overlapping and separated atoms. (F) Higher-order extrapolation comparing ProTEUS with MULTI-evolve (one-hot encoding FCNN). Models were trained on variants with up to K substitutions and tested on variants with more than K substitutions. Bars show mean Spearman correlation, points represent individual dataset evaluations, and error bars indicate SEM ($n = 13$, 15 and 12 for $K = 1$, 2 and 3, respectively). Datasets can contribute evaluations at more than one K. Asterisks indicate paired, one-sided Wilcoxon signed-rank tests for higher correlation with ProTEUS; *, $p < 0.05$. (G) In silico directed evolution across 17 multi-mutant datasets, comparing ProTEUS with EVOLVEpro, an additive model and MULTI-evolve (one-hot encoding FCNN). After an initial batch of 30 single mutants, each method selected 30 variants per round over five model-guided rounds. Lines and shading indicate mean ± SEM of normalized best fitness across datasets.

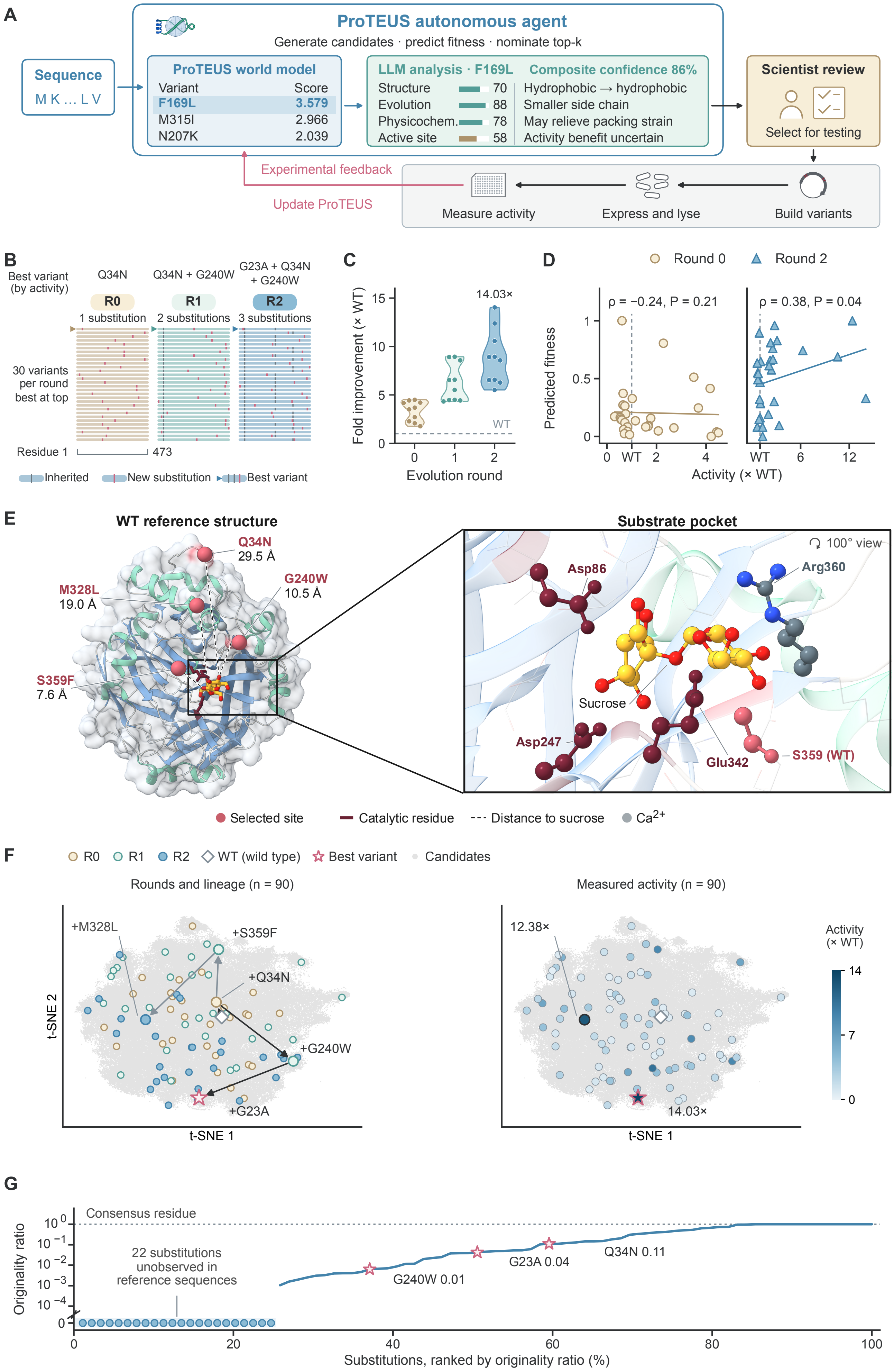
A
ProTEUS autonomous agent
Generate candidates · predict fitness · nominate top-k
Sequence
M K ... L V
ProTEUS world model
Variant Score
F169L 3.579
M315I 2.966
N207K 2.039
LLM analysis · F169L
Composite confidence 86%
Structure 70
Evolution 88
Physicochem. 78
Active site 58
Hydrophobic → hydrophobic
Smaller side chain
May relieve packing strain
Activity benefit uncertain
Scientist review
Select for testing
Experimental feedback
Update ProTEUS
Measure activity
Express and lyse
Build variants
B
Best variant (by activity)
Q34N
Q34N + G240W
G23A + Q34N + G240W
R0
R1
R2
1 substitution
2 substitutions
3 substitutions
30 variants per round best at top
Residue 1
473
Inherited
New substitution
Best variant
C
Fold improvement (× WT)
14.03×
WT
Evolution round
D
Round 0
Round 2
ρ = −0.24, P = 0.21
ρ = 0.38, P = 0.04
Predicted fitness
Activity (× WT)
E
WT reference structure
Q34N 29.5 Å
G240W 10.5 Å
M328L 19.0 Å
S359F 7.6 Å
Substrate pocket
100° view
Asp86
Arg360
Sucrose
Asp247
Glu342
S359 (WT)
Selected site
Catalytic residue
Distance to sucrose
Ca2+
F
R0
R1
R2
WT (wild type)
Best variant
Candidates
Rounds and lineage (n = 90)
Measured activity (n = 90)
+M328L
+S359F
+Q34N
+G240W
+G23A
t-SNE 2
t-SNE 1
12.38×
14.03×
Activity (× WT)
G
Consensus residue
22 substitutions unobserved in reference sequences
G240W 0.01
G23A 0.04
Q34N 0.11
Originality ratio
Substitutions, ranked by originality ratio (%)

# Figure 4

**Figure 4. Closed-loop evolution of levansucrase with ProTEUS**

(A) Experimental workflow linking variant nomination, construction, expression, sample preparation, activity measurement and feedback using an agentic version of ProTEUS for end to end closed-loop evolution. (B) Substitution map of 90 tested variants, with 30 variants per round ordered by decreasing activity. Marks distinguish inherited and newly introduced substitutions, and labels identify each round's best variant. (C) The cumulative top ten activity variants were shown for each round. Each point represents the mean of one variant. Violins show Gaussian kernel density estimates with bandwidth factor 0.35, restricted to the observed range. (D) Zero-shot scores in round 0 and composite scores in round 2 versus measured activity, including all 30 variants per batch. Scores are rescaled within each round. Lines show descriptive linear fits; Spearman correlations and P values are indicated. (E) Selected positions mapped onto WT SacB (PDB: 1OYG), with sucrose superposed from PDB 1PT2 (49). The enlarged pocket shows catalytic residues. (F) Shared t-SNE projection of candidate representations and all 90 assayed variants, colored by round and activity. Arrows trace parental lineages; the diamond denotes WT and the star denotes the final lead. (G) Originality ratios for 89 distinct substitutions, ranked from low to high. The ratio is the introduced-residue count divided by the largest amino-acid count at that position. Stars mark the final lead's substitutions; 22 unobserved substitutions are shown at zero below the axis break. Activities are relative to WT, and rounds are numbered from zero.

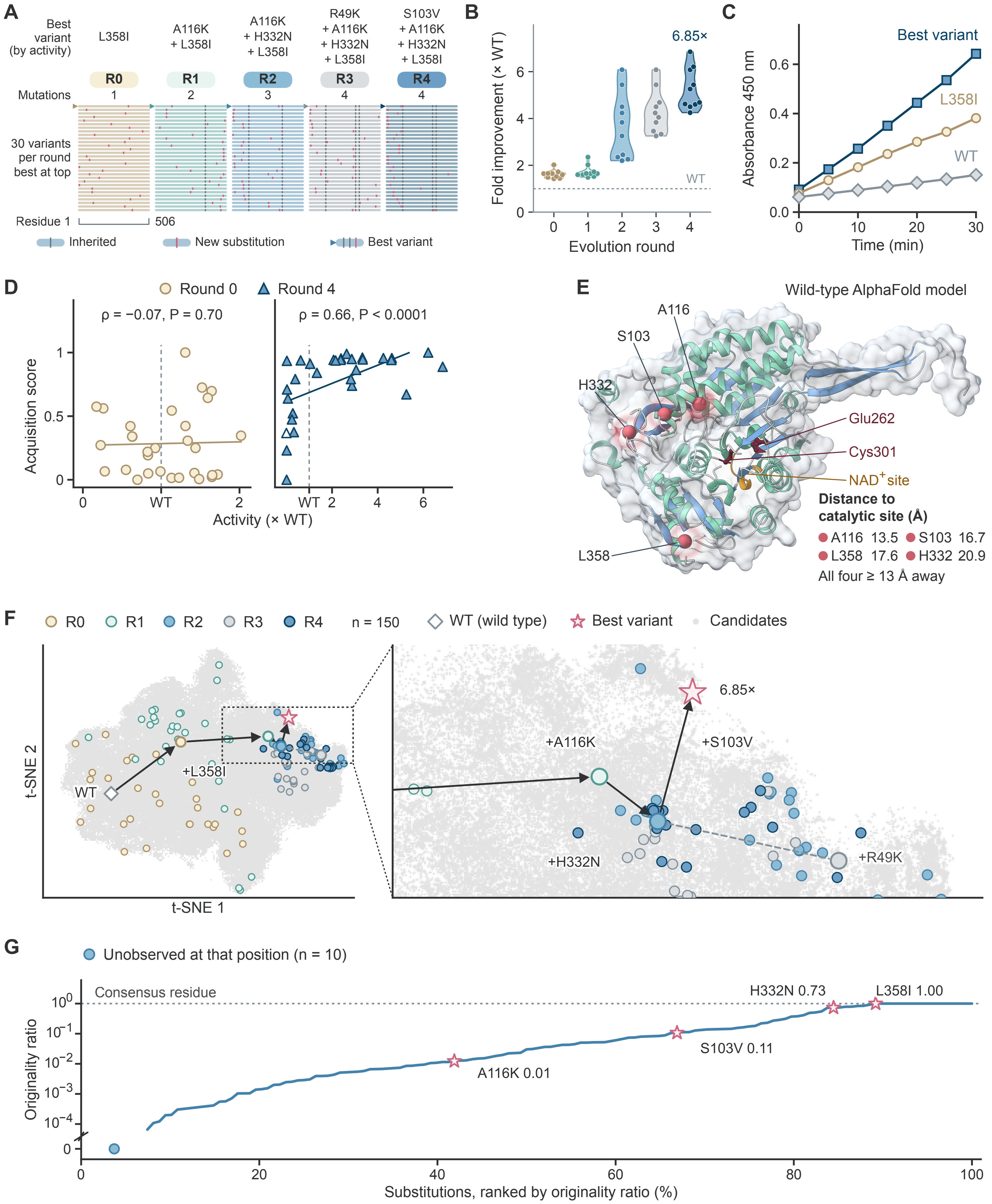

A
Best variant (by activity)
L358I
A116K + L358I
A116K + H332N + L358I
R49K + A116K + H332N + L358I
S103V + A116K + H332N + L358I
R0 R1 R2 R3 R4
Mutations 1 2 3 4 4
30 variants per round best at top
Residue 1 506
Inherited
New substitution
Best variant
B
Fold improvement (× WT)
6.85×
WT
Evolution round
C
Absorbance 450 nm
Best variant
L358I
WT
Time (min)
D
Round 0 Round 4
ρ = −0.07, P = 0.70
ρ = 0.66, P < 0.0001
Acquisition score
Activity (× WT)
E
Wild-type AlphaFold model
A116
S103
H332
Glu262
Cys301
NAD+ site
L358
Distance to catalytic site (Å)
A116 13.5 S103 16.7
L358 17.6 H332 20.9
All four ≥ 13 Å away
F
R0 R1 R2 R3 R4 n = 150 WT (wild type) Best variant Candidates
t-SNE 2
t-SNE 1
WT
+L358I
+A116K
+S103V
+H332N
+R49K
6.85×
G
Unobserved at that position (n = 10)
Consensus residue
Originality ratio
H332N 0.73
L358I 1.00
S103V 0.11
A116K 0.01
Substitutions, ranked by originality ratio (%)

# Figure 5

**Figure 5. Closed-loop evolution of aldehyde dehydrogenase with ProTEUS**

(A) Substitution map of 150 tested ALDH variants, with 30 variants per round ordered by decreasing activity. Marks distinguish inherited and newly introduced substitutions, and labels identify each round's best variant. Rounds 3 and 4 tested alternative fourth substitutions on A116K:H332N:L358I. (B) The cumulative top ten activity variants measured up to each round. Each point represents the mean of one variant. Violins show Gaussian kernel density estimates with bandwidth factor 0.35, restricted to the observed range. (C) Absorbance at 450 nm over 30 min of ALDH reaction assay using equal amounts of purified WT, L358I and the final lead ALDH candidate. Points and error bars show mean ± SD from three replicates per protein. (D) Zero-shot scores in round 0 and composite scores in round 4 versus measured activity, including all 30 variants per batch. Scores are rescaled within each round. Lines show descriptive linear fits; Spearman correlations and P values are indicated. (E) Positions altered in the final lead mapped onto AlphaFold reference model AF-P46368-F1 (*69*, *70*). Catalytic residues Glu262 and Cys301 and the annotated NAD+ site are highlighted; distances to the catalytic site are indicated. (F) t-SNE projection of candidate representations and 150 assayed variants colored by round, with an enlarged higher-order region. Arrows trace parental relationships, the diamond denotes WT, and the star denotes the final lead. (G) Originality ratios for 148 distinct substitutions, ranked from low to high. The ratio is the introduced-residue count divided by the largest amino-acid count at that position. Stars mark the final lead's substitutions; 10 substitutions were unobserved at their alignment positions. Screening activities are relative to WT, and rounds are numbered from zero.

**Supplementary Materials for**
**ProTEUS: A mutation-resolved world model for closed-loop protein evolution**

Yikun Zhang[1,5#,6#,*], Minghao Guo[2,*], Guangxue Xu[3], Peter Yichen Chen[4], Wojciech Matusik[2], Wengong Jin[5,6], Jingjing Chai[1], Kaiyi Jiang[1,7,†]

†Correspondence to: **kaiyi.jiang@princeton.edu**

# Materials and Methods

### Overview of the ProTEUS world model

ProTEUS represented protein variants in a joint space of sequence, evolutionary and structural information. Mutation actions specified substitutions applied to WT or selected parental sequences, and the model predicted the fitness of the resulting candidates. Deep mutational scanning (DMS) pretraining established the shared representation, while target-specific measurements supported supervised fine-tuning or iterative prediction updates. A nomination strategy connected these predictions to successive experimental batches, with returned measurements informing the next round of candidate selection.

### Deep mutational scanning datasets for pretraining and evaluation

We assembled the pretraining data from 217 substitution deep mutational scanning (DMS) assays in ProteinGym (*29*). To separate pretraining from downstream evaluation, we compared assay wild-type (WT) sequences with those of 38 reference sequences from held-out datasets reserved for downstream evaluation (fig. S1A and fig. S1B). Screening used exact-sequence, bidirectional-substring and UniProt-identifier matching, followed by MMseqs2 (*71*) homology searches. We excluded pretraining assays with at least 30% sequence identity and 0.5 query coverage to an evaluation protein. After removal of 35 assays, the corpus contained 182 assays representing 161 distinct WT sequences and 160 UniProt entries. The pretraining assays covered proteins of 37 – 3,423 amino acids (median, 224) and contained 63 – 496,137 measured variants per assay (median, 2,238). The corpus comprised 65 stability, 61 organismal-fitness, 32 activity, 15 expression and nine binding assays, with 80 human, 41 prokaryotic, 31 other eukaryotic and 30 viral assays. We oriented experimental scores so that larger values indicated better measured performance and normalized them within each assay. Variants with unresolved sequences or reported WT residues inconsistent with the reference sequence were removed. Variants from the 182 assays were pooled and divided into training, validation and test sets in a 3:1:1 ratio. The test set therefore assessed prediction on held-out variants from pretraining assays (Fig. 2B). Each architecture comparison used the same held-out variants for all models within that comparison (Fig. 2D). Zero-shot prediction and target-specific supervised fine-tuning were evaluated on 19 assays excluded from pretraining. The zero-shot benchmark retained the available mixture of single- and multi-substitution variants in each assay (Fig. 2E); the first-round nomination benchmark used 13 single-substitution assays (Fig. 2F).

For higher-order analyses, we assembled 17 phenotype-specific DMS datasets covering 12 proteins. These comprised 14 datasets from CombinGym (*46*) and combinatorial datasets for AAV, GFP and TrpB. Different phenotypes measured for the same protein were analyzed as separate datasets. The higher-order extrapolation benchmark used 15 datasets and the iterative multi-mutant benchmark used all 17 datasets.

### Mutation-resolved multimodal architecture and pretraining

Each variant state was encoded relative to the WT sequence for its assay using aligned sequence, MSA and structural inputs. Inputs contained at most 512 residues. For longer proteins, all three streams used a common mutation-containing window and residue offset. The mutation mask, Mmut,i, was one at substituted positions and zero elsewhere, where i indexes aligned residues.

For each variant, we extracted residue-level features from sequence, MSA, and structure encoders. Sequence features were taken from the final layer of ESM-2 3B (*12*) after removing beginning- and end-of-sequence tokens, yielding a 2,560-dimensional vector at each residue. The MSA stream used the 768-dimensional query-row representation from MSA Transformer (*15*). We used ESM3 (*21*) as the structure encoder.

For each WT protein, homologous sequences were retrieved and aligned to the target. Redundant sequences were removed with hhfilter (*72*), and each alignment was limited to 128 sequences for the primary model.

A modality-specific factorized linear adapter mapped each native feature dimension through a 64-dimensional bottleneck to the shared width of 768. For each aligned residue $i$, the input feature $f_{m,i}$ was projected as

$$H_{m,i} = W_{m,2}\left(W_{m,1}f_{m,i} + b_{m,1}\right) + b_{m,2} + e_m + p_{m,i}. \qquad (1)$$

Here, the modality subscript $m$ denotes sequence (seq), MSA, or structure (str). The adapter contains two linear projections with weights $W$ and biases $b$, without an intervening nonlinearity. Each projected feature received a learned modality embedding, $e_m$. The positional term, $p_{m,i}$, was zero for sequence and MSA features, which retained positional information from their source encoders. The structure stream received a fixed one-dimensional sine-cosine position encoding (*73*). The aligned representations were summed at each residue to form the MutationFormer input:

$$H_{\mathrm{fused},i} = H_{\mathrm{seq},i} + H_{\mathrm{MSA},i} + H_{\mathrm{str},i}. \qquad (2)$$

A mutation mask marked every substituted residue, allowing the same encoder to represent single and multi-mutant variants.

MutationFormer comprised 12 pre-normalized transformer blocks with hidden width 768 and 12 attention heads. Each head had 64 dimensions. Within every head and block, the attention logit between query position $i$ and key position $j$ was

$$A_{ij} = \frac{Q_i^{\top} K_j}{\sqrt{d_{\text{head}}}} + b_{\text{mut}} M_{\text{mut},j} + M_{\text{pad},j}. \qquad (3)$$

Here, $Q$, $K$, and $V$ are the query, key, and value representations, and $d_{\text{head}} = 64$. The mutation-bias strength, $b_{\text{mut}}$, was fixed at 10. The padding mask, $M_{\text{pad},j}$, was zero for valid keys and negative infinity for padded positions. Adding the mutation bias at key positions increased attention to substituted residues from every query. Attention weights and head outputs were

$$\text{Attn}_{ij} = \frac{\exp(A_{ij})}{\sum_{j'} \exp\left(A_{ij'}\right)}, \quad O_i = \sum_j \text{Attn}_{ij} V_j. \qquad (4)$$

Each block applied mutation-biased multi-head attention (MHA), followed by a SwiGLU feed-forward network (FFN) (*74*), with separate pre-layer normalizations (LN) and residual connections. Writing $H_{\text{in}}$, $H_{\text{attn}}$, and $H_{\text{out}}$ for the block input, post-attention representation, and block output, respectively,

$$H_{\text{attn}} = H_{\text{in}} + \text{MHA}_{\text{mut}}[\text{LN}_1(H_{\text{in}})], \quad H_{\text{out}} = H_{\text{attn}} + \text{FFN}[\text{LN}_2(H_{\text{attn}})]. \qquad (5)$$

$$\text{FFN}(H) = W_2[\text{SiLU}(W_1 H) \odot W_3 H]. \qquad (6)$$

The feed-forward branch used SiLU activation and elementwise multiplication to gate its intermediate features. The query, key, value, attention-output, and feed-forward projections were bias-free. The gated feed-forward branch mapped 768 dimensions to 2,048 dimensions and back to 768.

The normalized MutationFormer embedding, $H_{\text{reg}}$, was used to predict variant fitness through a neural fitness head:

$$y_{\text{pred}} = w_{\text{out}}^{\top} \text{LN}_{\text{head}}\left[\text{GELU}\left(W_{\text{head}} H_{\text{reg}} + b_{\text{head}}\right)\right] + b_{\text{out}}. \qquad (7)$$

Here, $y_{\text{pred}}$ is predicted fitness, and the head contains trainable weights and biases. ProTEUS base model was trained jointly on variants from the 182 DMS assays to predict assay-normalized experimental fitness. We minimized mean squared error between predictions and assay-normalized measurements, denoted by $y_{\text{meas}}$. For a minibatch containing $B$ variants, indexed by $n$, the objective was

$$\mathcal{L}_{\text{fitness}} = \frac{1}{B}\sum_{n=1}^{B}\left(y_{\text{pred},n} - y_{\text{meas},n}\right)^2. \qquad (8)$$

Trainable components and the distinctions between pretraining, target-specific neural fine-tuning and model updating during iterative evolution are specified in table S2.

We evaluated ProTEUS base model and 13 baselines on the test set held out from the 182-assay pretraining data (Figure 2B). Spearman rank correlation measured agreement between predicted and measured variant rankings, and MSE measured error in normalized fitness.

To assess data scaling, we sampled 1.25%, 2.5%, 5%, 10%, 20%, 40%, 80% and 100% of the original training set and pretrained one model at each fraction using the same procedure described above. The validation and test sets were held fixed across fractions, and performance was evaluated by test-set MSE and Spearman correlation (Fig. 2C). We fitted an empirical power law by unweighted linear regression of log-transformed MSE on log-transformed training-data fraction, with an intercept, across all eight fractions. Goodness of fit was evaluated using $R^2$ in log space.

We evaluated mutation bias, input modalities, and MSA sampling in separate ablation experiments (Figure 2D). Setting the mutation bias to zero removed the attention bias toward substituted positions while retaining the rest of the architecture. We tested bias strengths of 0, 5, 10, 15, and 20. The complete model used a bias of 10, which gave the highest Spearman correlation. Modality ablations compared different combinations of sequence, MSA, and structural inputs. For the MSA-sampling comparison, we either removed redundant sequences with hhfilter or sampled 128 sequences from the unfiltered alignment according to sequence weights. Models within each comparison were evaluated on the same held-out variants.

### Zero-shot variant ranking and initial nomination

Zero-shot denotes scoring without fitness measurements from the evaluated target. Sequence- and MSA-based mutant-versus-WT log-likelihood differences were summed across substituted positions, and MSA logits were averaged over independently sampled alignments (*13*). Modality scores were combined using weights derived from the modality embeddings learned during DMS pretraining:

$$S_{\text{ZS}} = \sum_{m} w_m S_m, \quad w_m = \frac{\text{mean}(|e_m|)}{\sum_{m'} \text{mean}(|e_{m'}|)}. \qquad (9)$$

The sums included only the modalities used for a given score. For the 19-assay zero-shot benchmark, ProTEUS base model combined the sequence and MSA terms (Fig. 2E). Spearman correlation was calculated separately for each assay and macro-averaged with equal weight across assays. This score was distinct from the output of the neural fitness head used for supervised prediction.

For first-round nomination, we compared ProTEUS base model with a rank-fusion baseline adapted from the four zero-shot scores used by MULTI-evolve (*1*) across 13 single-substitution MaveDB (*33*) assays (Fig. 2F; fig. S2B). The four combinatorial Holdout4 datasets (AAV, GB1, GFP and TrpB) and two MaveDB assays containing multi-substitution variants (urn_mavedb_00000105-a-1 and urn_mavedb_00000665-a-1) were outside this comparison.

The adapted MULTI-evolve baseline used sequence- and structure-conditioned mutant-versus-WT scores and their substitution-type-standardized counterparts. The sequence-conditioned score combined five ESM-1v models (*13*) with ESM-2 3B (*12*); the structure-conditioned score used ESM-IF1 (*19*). We averaged within-assay ranks across the four scores to evaluate each method at the same total nomination budget. This adaptation differs from the published procedure, which nominates candidates independently from each score and takes their union. Before nomination, only the highest-ranked substitution at each residue position was retained. Batch sizes were 5, 10, 15, 24, 50 and 100 variants. A hit had an assay-standardized DMS score of at least zero, corresponding to measured fitness at or above the assay mean.

### Target-specific learning and single-mutant evolution

We evaluated target-specific supervised fine-tuning using measured variants from assays excluded from pretraining (Fig. 2G). Trainable parameters were initialized from DMS pretraining and optimized against target-assay measurements using the same MSE objective. The trainable components are specified in table S2.

Fivefold cross-validation was performed independently for each assay, with four folds used for fine-tuning and the remaining fold for evaluation. Fold assignments were identical across methods, and metrics were averaged over folds within each assay.

Single-mutant evolution was evaluated over five rounds on CALM1, GAL4, HRAS, MAPK1, TPK1, TPMT, UBE2I, UBI4, beta-lactamase and HaeIII methyltransferase datasets (Fig. 2H). ProTEUS and EVOLVEpro (*3*) each selected 30 variants per round from the same candidate library for each dataset. Accumulated measurements were used to update fitness predictions before each subsequent round of selection.

At each round, we calculated the fraction of nominated variants belonging to the top 5%, 10% or 20% of the complete DMS dataset, with Precision@10 additionally compared across 14 datasets at batch sizes of 30, 50 and 100 variants per round (fig. S3A–C). The best fitness recovered by that round was normalized to the observed fitness range of the dataset. Round-wise trajectories report the mean normalized best fitness across datasets.

### Physics-constrained all-atom structure prediction

We developed an all-atom predictor to generate variant-specific structures for multi-mutant candidates (Fig. 3A–E). The predictor retained the atom-attention encoder, MSA stack, PairFormer and diffusion coordinate module of Boltz-1 and was initialized from its publicly released weights (*43*). We fine-tuned the predictor with the denoising loss evaluated on projected coordinates, as described below (*32*). At inference, two denoising steps used molecular conditioning features to estimate denoised coordinates and update the noisy coordinates:

$$\bar{X}^{(s)} = D\left(\tilde{X}^{(s)}, \hat{t}_s; H_{\mathrm{cond}}\right), \quad X^{(s+1)} = \tilde{X}^{(s)} + \eta_s (t_{s+1} - \hat{t}_s) \frac{\tilde{X}^{(s)} - \bar{X}^{(s)}}{\hat{t}_s}. \tag{10}$$

Here, $s = 0,1$ indexes the two steps; $D$ is the denoiser and $H_{\mathrm{cond}}$ contains molecular conditioning features. $\tilde{X}^{(s)}$ denotes centered, randomly augmented coordinates after noise injection, $\bar{X}^{(s)}$ is the denoised estimate, $\hat{t}_s$ is the noise level after inflation, $t_{s+1}$ is the scheduled next noise level and $\eta_s$ is the step factor. We retained the original noise parameterization and denoising objective and used the two-step schedule at inference (*32*, *75*). Coordinates after the second update were corrected by differentiable projection.

The projection enforced seven constraint families: nonbonded inter-chain steric separation, tetrahedral chirality, double-bond E/Z stereochemistry, carbon double-bond planarity, ligand internal-distance geometry, separation of symmetric chains and covalent bonds spanning separate chains (table S3). Each was expressed as a hinge residual that vanished within its feasible region. Physical validity in the reported evaluation required satisfaction of all seven specified constraint families to numerical tolerance.

Projection balanced deviations from the predicted coordinates against violations of the physical constraints. Given predicted coordinates $X_{\mathrm{pred}}$ and constraint residuals $C(X)$, we used a quadratic exterior penalty (*76*):

$$X_{\mathrm{proj}} = \underset{X}{\mathrm{argmin}} \left[\frac{1}{2}\left\|X - X_{\mathrm{pred}}\right\|_F^2 + E_{\mathrm{pen}}(X)\right], \quad E_{\mathrm{pen}}(X) = \frac{1}{2} C(X)^{\mathsf{T}} A_{\mathrm{pen}}^{-1} C(X). \tag{11}$$

The diagonal matrix $A_{\mathrm{pen}}$ contained coefficients of $10^{-7}$ or $10^{-6}$, depending on the constraint family. Their small positive values imposed strong penalties on constraint violations. We introduced multipliers $\lambda = -A_{\mathrm{pen}}^{-1} C(X)$ and used in-place Gauss–Seidel updates instead of assembling the full linear system (*77*). For constraint $j$ with diagonal coefficient $a_{\mathrm{pen},j}$, the update was

$$\Delta\lambda_j = \frac{-C_j(X) - a_{\mathrm{pen},j}\lambda_j}{\left\|\nabla C_j(X)\right\|_2^2 + a_{\mathrm{pen},j}}, \qquad X \leftarrow X + \nabla C_j(X)^{\mathsf{T}} \Delta\lambda_j, \qquad \lambda_j \leftarrow \lambda_j + \Delta\lambda_j. \tag{12}$$

Each update modified only atoms involved in the current constraint, and subsequent constraints used the updated coordinates. We applied 20 complete sweeps in both training and inference. Constraint groups with no shared atoms were processed in parallel on the GPU, giving linear work in the number of local constraints per sweep.

The projection $P_{\mathrm{proj}}$ was included in the training graph, and the denoising loss compared projected coordinates with the reference structure $X_{\mathrm{ref}}$:

$$\mathcal{L}_{\mathrm{str}} = \mathbb{E}\left[\ell_{\mathrm{denoise}}\left(P_{\mathrm{proj}}\left(D(X_{\mathrm{noisy}}, t; H_{\mathrm{cond}})\right), X_{\mathrm{ref}}\right)\right]. \tag{13}$$

$X_{\mathrm{noisy}}$ denotes the noisy training coordinates and $t$ the training noise level. We computed gradients through the projection by implicitly differentiating its optimality condition, avoiding the need to store intermediate iterates for backpropagation (*32*, *78*). With penalty Hessian $H_{\mathrm{pen}}$, identity matrix $I$ and adjoint variable $z$, the backward pass solved

$$[H_{\mathrm{pen}}(X_{\mathrm{proj}}) + I]^{\mathrm{T}} z = \left(\frac{\partial \mathcal{L}_{\mathrm{str}}}{\partial X_{\mathrm{proj}}}\right)^{\mathrm{T}}, \qquad \frac{\partial \mathcal{L}_{\mathrm{str}}}{\partial X_{\mathrm{pred}}} = z^{\mathrm{T}}. \tag{14}$$

The adjoint system was solved by conjugate gradients with a tolerance of $10^{-4}$ and at most 1,000 iterations. Forward and backward projection were implemented as a custom PyTorch autograd layer (*79*) in Warp (*80*), without host–device coordinate transfer. The model was fine-tuned on approximately 180,000 structures from the RCSB Protein Data Bank (*63*), using eight NVIDIA H100 GPUs.

Structural evaluation used six biomolecular-complex benchmarks (*32*). We assessed complex local distance difference test (LDDT) (*44*), benchmark-specific accuracy metrics and the seven physical-validity criteria (Fig. 3C). Comparators included AlphaFold 3 (*42*), Boltz-1 (*43*), Protenix (*66*), Boltz-1-Steering (*43*), Boltz-2 (*81*) and five- and ten-step Protenix-Mini (*75*). Within each structural benchmark, metric values were first averaged across generated samples for each protein. Benchmark means and 95% confidence intervals were then calculated across these protein-level means. Reported ± values denote the half-width of the 95% confidence interval.

Structural evaluation and the original CASP15 runtime benchmarks were conducted on NVIDIA A6000 GPUs (*32*). The reported inference times included diffusion and physical-validity enforcement where applicable. Runtime comparisons additionally included AlphaFold 3 mean runtimes for the same five atom-count bins, spanning fewer than 1,000 to at least 4,000 atoms (Fig. 3D). For each bin, speedup was calculated as the comparator mean runtime divided by the mean runtime of the two-step predictor. Median speedups were calculated across the five bin-specific ratios.

### Higher-order fitness prediction from lower-order measurements

We evaluated extrapolation to higher-order variants using a K-vs-rest design (*46*) with $K = 1$, $2$ or $3$ (Fig. 3F). For each dataset and $K$, the training set contained WT, when available, and measured variants with at most $K$ substitutions. The test set contained measured variants with more than $K$ substitutions. For each dataset–K combination, ProTEUS was evaluated after supervised neural fine-tuning on the lower-order training variants. Spearman correlation was calculated on the corresponding higher-order test set.

The analysis comprised 40 comparisons across 15 datasets representing 11 proteins: 13 comparisons for $K = 1$, 15 for $K = 2$ and 12 for $K = 3$. Seven of the 47 available comparisons were excluded. CR6261-H9 at $K = 1$ had no variation in training fitness. For SpCas9 and CR6261-H1, comparisons at $K = 1$, $2$ and $3$ were excluded because the single-mutant training sets were small and most substitutions in the test variants occurred at positions absent from training.

We evaluated MULTI-evolve using the one-hot sequence encoding and two-layer fully connected network configuration in its public implementation. The model was trained separately for each dataset–K combination.

### In silico evolution of higher-order variants

Multi-mutant evolution was evaluated on 17 DMS datasets using ProTEUS, EVOLVEpro, an additive model and the MULTI-evolve one-hot predictor (Fig. 3G). Each simulation comprised an initial batch of 30 single mutants and five model-guided rounds of 30 variants, giving a total budget of 180 variants per dataset. Each method used the same candidate library and nomination budget for a given dataset. After each batch, accumulated measurements were used to update fitness predictions and rank untested candidates.

In the progressively expanding benchmark schedule, the first model-guided round admitted double mutants, and each subsequent round increased the maximum permitted number of substitutions by one. Untested lower-order variants remained eligible, whereas previously tested variants were excluded. Separate comparisons of nomination strategies used up to five model-guided rounds, with datasets contributing while eligible untested candidates remained (fig. S4; table S4).

The additive model used binary indicators of constituent substitutions and ridge regression; regularization was selected by up to fivefold cross-validation over 12 values from $10^{-3}$ to $10^{4}$. The MULTI-evolve neural-network baseline was retrained after each round and selected the variants with the highest predicted fitness.

### Closed-loop variant nomination and experimental feedback

ProTEUS linked candidate generation, virtual fitness evaluation, batch nomination and experimental feedback within a single workflow (Fig. 1A–C). Mutation actions generated candidate sequences from WT or selected parental backgrounds within the permitted mutation depth. The platform prepared sequence, MSA and structural inputs, evaluated candidates and nominated variants for laboratory testing. Returned measurements were added to the target-specific training data to update fitness predictions before the next round of selection.

The initial experimental batch contained 30 single mutants selected by zero-shot ranking. Their measured activities were used to update fitness predictions for subsequent nominations. The model-update procedure and its relationship to supervised pretraining and fine-tuning are described in table S2.

Subsequent experimental rounds could retain the current maximum number of substitutions or admit variants carrying one additional substitution. Previously tested variants were excluded from subsequent batches.

After the initial zero-shot batch, nomination combined within-library zero-shot ranks, updated fitness-prediction ranks and fitness ranks derived from all available experimental measurements:

$$S_{\mathrm{nom}} = w_{\mathrm{ZS}} R_{\mathrm{ZS}} + w_{\mathrm{RF}} R_{\mathrm{RF}} + w_{\mathrm{CF}} R_{\mathrm{CF}}. \quad (15)$$

The ranks $R_{\mathrm{ZS}}$, $R_{\mathrm{RF}}$ and $R_{\mathrm{CF}}$ corresponded to these three components, respectively, and were oriented so that larger values indicated higher priority. The weights were nonnegative. Eligible, untested candidates were ranked under each weighting configuration, and experimental batches were assembled from the resulting shortlists. Simulation nomination strategies and laboratory evolution batches are summarized in tables S4 and S5, respectively. Measurements from each selected batch informed the next cycle of prediction and nomination.

For the ALDH campaign, the random-forest predictor used 768-dimensional mean-pooled encoder representations and was refitted using all accumulated experimental measurements.

## Mutant library construction and expression

Individual amino acid substitutions were introduced into *Bacillus subtilis* levansucrase and *Cupriavidus pinatubonensis* ALDH by inverse PCR using Q5 High-Fidelity DNA Polymerase (NEB). Back-to-back primers were centered on each target codon, with 15–25 bp template homology on each side and melting temperatures of 60–72 °C. Methylated templates were digested with DpnI, and products were circularized with NEBuilder HiFi DNA Assembly Master Mix (NEB) at 50 °C for 15–60 min. Assemblies were transformed into NEB Stable competent cells. Two to three colonies per mutant were picked and miniprepped. Sequencing confirmed the intended substitution and absence of secondary mutations across the coding region. One verified clone per mutant was retransformed into NEB Express cells for expression. Mutants were processed in batches of approximately 30 in 96-deep-well plates, alongside three WT replicates and three empty-vector controls per plate. Starter cultures contained 1 mL LB medium with antibiotic and were grown overnight at 37 °C and 250 rpm. Cultures were diluted 1:30 into 1.5 mL fresh medium and grown to $OD_{600}$ 0.4–0.6; the $OD_{600}$ at induction was recorded for each well. Expression was induced with 0.5 mM IPTG, and cultures were shifted to 25 °C for 4–6 h. Final $OD_{600}$ was recorded before harvest by centrifugation at 3,000–4,000 × g for 15 min at 4 °C.

## Activity assays for experimental screening

Levansucrase activity was measured in CTAB-permeabilized cells. Pellets were resuspended in McIlvaine's buffer (pH 6.0), normalized to the same optical density and treated with 0.1% CTAB for 10 min at 900 rpm. Sucrose was added to a final concentration of 100 mM, and reactions proceeded at 37 °C for 5 min. Reactions were stopped by dilution into 200 mM Tris (pH 8.3) and heat inactivation at 96 °C for 5 min. Released glucose was quantified with the Abcam Glucose Assay Kit (ab65333) as the activity readout.

ALDH activity was measured in clarified crude lysate. Pellets were resuspended in ice-cold kit assay buffer containing fresh protease inhibitors and sonicated on ice at 20–30% amplitude. Sonication comprised three cycles of 10 s on and 10 s off. Lysates were clarified by centrifugation at 12,000–15,000 × g for 10 min at 4 °C. Lysate volumes of 10–20 µL were adjusted to 50 µL with assay buffer and combined with reaction mix from the Abcam ALDH Activity Assay Kit (ab155893). NADH production was monitored using the coupled absorbance readout at 450 nm and 37 °C. Activity was quantified from the slope of the initial linear region of the kinetic curve recorded over 20–30 min.

### Enzyme quantification and concentration-normalized ALDH activity

Both enzymes carried a C-terminal His tag. Both enzyme concentrations were quantified using a His-tag enzyme-linked immunosorbent assay (ELISA) kit (GenScript, L00436), following the manufacturer's protocol. Standard curves were fitted in GraphPad Prism using a four-parameter logistic model, with $R^2 > 0.9$ required for acceptance. Variant concentrations were estimated from the fitted curves. Measured ALDH activity was divided by enzyme concentration to obtain concentration-normalized activity. Enzyme concentration and concentration-normalized activity were each expressed relative to the corresponding WT value.

### Data and model implementation

#### *Pretraining data and similarity analyses*

Variant sequences in the ProteinGym (*29*) deep mutational scanning (DMS) corpus were reconstructed from reported substitutions and the assay-specific wild-type (WT) sequence.

Sequence alignments from MMseqs2 (*71*) were summarized by identity, query coverage and aligned length. Structural alignments from Foldseek/TMalign (*82*, *83*) were ranked by query-normalized TM-score, retaining the best match for each reserved entry. The structural analysis included 181 available pretraining structures and 38 reserved structures. Sequence and structural summaries are shown in Fig. S1.

#### *Model training and target-specific updates*

The ESM-2 (*12*), MSA Transformer (*15*) and ESM3 (*21*) source encoders (Table S1) remained fixed. DMS pretraining jointly optimized the multimodal representation model and neural fitness head (Table S2). Target-specific neural fine-tuning updated the same trainable components. The neural fitness head used the final layer-normalized register-token embedding. This learned token was prepended to the residue sequence with a mutation-mask value of zero.

During iterative evolution, the pretrained representation model provided fixed inputs to a target-specific random forest. The random forest was refitted after each round using all accumulated experimental measurements.

*Geometric and stereochemical constraints*

Table S3 specifies the geometric and stereochemical bounds enforced by Gauss–Seidel projection (*32*). The predictor architecture, training procedure and projection equations are described in the section “Physics-constrained all-atom structure prediction”.

## Computational benchmarks

### *Single-mutant evolution and experimental budgets*

The single-mutant benchmark used one ProTEUS run (Figure 2H). EVOLVEpro (*3*) used mean-pooled ESM-2 15B embeddings and a 100-tree random forest, with results averaged over 10 runs.

The expanded analysis added AAV, GB1, GFP and TrpB to the 10 single-mutant datasets. Dataset-level Precision@10 is shown for batches of 30, 50 and 100 variants per round (Fig. S3). Best-fitness summaries cover batch sizes of 30–100 in increments of 10. Normalized best-fitness values were summarized without additional rescaling.

### *Higher-order prediction and multi-mutant evolution*

Higher-order predictions were compared with the MULTI-evolve supervised predictor (*1*) using one-sided paired Wilcoxon signed-rank tests for higher Spearman correlation with ProTEUS within each K group (Figure 3F). Here, K is the maximum number of substitutions in training variants. Unadjusted P values were 0.0133, 0.0108 and 0.0386 for K = 1, 2 and 3, respectively.

Multi-mutant evolution simulations used seed 42 for ProTEUS and seeds 1, 2, 42, 2025 and 2026 for the comparators (Figure 3G). Best-fitness values were min–max normalized. Comparator results were averaged over seeds within each dataset before calculating the across-dataset mean and standard error of the mean (SEM). No further rescaling was applied across methods or rounds.

### *In silico comparison of nomination strategies*

We compared five nomination strategies using fixed ProTEUS representations and a random-forest predictor updated after each batch (Fig. S4; Table S4). The panel comprised 13 CombinGym (*46*) phenotype datasets and AAV, GB1 and TrpB. For each dataset, the cumulative mean DMS score was divided by the largest cumulative mean obtained across all evaluated strategies and rounds. Across-dataset summaries report the mean and sample standard deviation (SD). The numbers of contributing datasets were 15, 16, 13, 11 and 11 in model-guided rounds 1–5, respectively.

## Laboratory nomination and batch selection

Both levansucrase and aldehyde dehydrogenase (ALDH) campaigns began with zero-shot variant ranking. Subsequent nominations combined zero-shot ranks, updated random-forest prediction ranks and fitness ranks derived from all available experimental measurements. Each experimental batch contained 30 variants, and round labels begin at zero. Table S5 lists the parental backgrounds, substitution counts and best variants for each batch.

For levansucrase, candidate nominations drew on shortlists generated under different weighting configurations. The final batch retained two parental backgrounds, comprising 18 triple mutants derived from Q34N:G240W and 12 derived from Q34N:S359F.

For ALDH, five weighting configurations were evaluated for zero-shot ranks and fitness ranks derived from experimental measurements. The respective weight pairs were (0, 0.2), (0, 0.4), (0, 0.6), (0.2, 0.2) and (0.2, 0.4). The random-forest weight was one minus the sum of these two weights. Each configuration produced a shortlist of 30 variants. Candidates were prioritized by the number of shortlists in which they appeared, with ties ordered by their best rank across the shortlists. The top 30 candidates were selected for experimental testing.

## Analysis of laboratory evolution campaigns

### *Score–activity associations*

Score–activity analyses used the zero-shot ranking in round 0. In subsequent rounds, composite scores combined normalized zero-shot ranks, random-forest prediction ranks and fitness ranks derived from all available experimental measurements, with respective weights of 0.2, 0.6 and 0.2. Scores were rescaled to 0–1 within each round, and Spearman correlations included all 30 assayed variants.

For levansucrase, composite scores were analyzed in rounds 1 and 2.

For ALDH, the reference sequence was UniProt P46368. All nine nonpositive activity measurements were retained in the score–activity analysis. Cumulative best-activity and top-ten summaries included all variants tested up to each round.

### *Protein representations and geometric analyses*

For levansucrase, sucrose from PDB 1PT2 was transferred to WT SacB (PDB 1OYG) after Cα-based structural superposition. Residue–substrate distances were the minimum heavy-atom separations between each WT residue and the superposed sucrose. For ALDH, substitutions were mapped to WT model AF-P46368-F1. Distances to the catalytic site were the minimum heavy-atom separations between each WT residue and either Glu262 or Cys301.

Quantitative representation analyses used the original 768-dimensional vectors. Candidate libraries and assayed variants were visualized using t-distributed stochastic neighbor embedding (t-SNE). Parent–child relationships were defined by shared and newly introduced substitutions.

For levansucrase, mean-pooled ProTEUS representations were available for WT and 90 assayed variants. The candidate background contained 366,836 library entries across single-, double- and triple-mutant libraries. The t-SNE map was fitted to 20,000 points balanced across these libraries, and the remaining entries were projected into the fitted map. Within each final-round parental background, we compared cosine distances among pairs at amino acid Hamming distance two. Principal component analysis (PCA) of mean-centered representations determined the number of components explaining 90% of the variance; the comparator encoded the final amino acid at each variable position by one-hot encoding (Table S9).

For measured levansucrase parent–child pairs, the mutational displacement was the child representation minus the parental representation. We compared the two displacement vectors for I369H, the only added substitution tested on both final-round parents.

For ALDH, all variants were encoded with the same pretrained 12-layer checkpoint. Final encoder states were averaged over valid positions, including the register token and excluding padding. The candidate background contained 9,615 single-substitution, 153,402 double-substitution, 47,873 triple-substitution and 143,254 four-substitution library entries, totaling 354,144. Counts describe library entries rather than deduplicated sequences. The 150 assayed variants and WT were matched to their representations by substitution identity.

The ALDH projection used a subset stratified across the four candidate libraries, reduced to 50 dimensions by PCA. We fitted t-SNE with openTSNE (*51*, *84*), using perplexity 50, PCA initialization and random seed 42. Remaining library entries, assayed variants and WT were transformed using the same PCA basis and fitted t-SNE map.

### *Local activity association*

We compared absolute activity differences between embedding neighbors with those between sequence-distance-matched variants. Neighbors were defined by cosine distance in the original representation space, excluding the anchor itself. For each directed anchor–neighbor pair, the reference set contained all variants in the same analysis group at the same amino acid Hamming distance from the anchor, including the selected neighbor. Hamming distance counted residue positions that differed between variants. Differences were averaged within anchors before aggregation across variants.

The statistic was the matched-reference mean absolute activity difference (MAD) minus the neighbor MAD; relative reduction was one minus the ratio of neighbor MAD to reference MAD. Activity labels were permuted 19,999 times within analysis groups, keeping neighborhoods and sequence-distance matrices fixed. Two-sided P values were twice the smaller tail probability, with a plus-one correction and a maximum of 1. Variants were the permutation units, and all recorded activities were retained.

For levansucrase, the analysis included the 30 triple mutants from the final round, comprising 18 descendants of Q34N:G240W and 12 of Q34N:S359F. Neighbors, reference sets and label permutations were restricted to each parental background. All 30 anchors were weighted equally. We evaluated three, five and ten neighbors, using random seed 20261028.

For ALDH, the analysis included the 60 four-substitution variants from rounds 3 and 4. Neighbors, reference sets and label permutations were restricted to each experimental batch. Differences were averaged across variants within each batch and then equally across the two batches. Five neighbors defined the primary analysis; three and ten neighbors were used for sensitivity analyses (Table S6). The random seed was 20261019.

### *Retrospective activity prediction*

Representation-based predictors were evaluated retrospectively, separately from the variant-nomination procedure. Performance was summarized by Spearman correlation and mean absolute error (MAE).

For levansucrase, ridge-regression and random-forest readouts compared embeddings with substitution one-hot features, training-mean predictions and measured-parent activities. Evaluation used round 0 to predict round 1, rounds 0–1 to predict round 2, and nested leave-one-out prediction within round 2. Model selection used training data only.

For ALDH, the 30 round-3 variants formed the training set and the 30 round-4 variants formed the test set. All variants shared the same parental background and substitution count. Predictions were the mean activities of the five nearest training variants by cosine distance, without tuning. Comparators used the training-set mean, training-set median or Hamming-distance neighbors. The Hamming-distance comparator assigned equal weights to all variants tied at the fifth-neighbor distance.

Uncertainty in ALDH prediction was assessed using 10,000 paired bootstrap resamples of the 30 test variants, keeping training data fixed, to obtain percentile 95% confidence intervals. A two-sided correlation test used 19,999 permutations of test activity labels. Intervals describe test-variant uncertainty conditional on the observed training data and activity measurements.

### *Alignment-based substitution frequencies*

At each substituted position, we counted occurrences of the introduced amino acid among homologous sequences. The originality ratio was the introduced-residue count divided by the highest amino acid count at the same position. A ratio of 1 denoted a consensus residue, and 0 denoted a residue not observed at that position in the alignment. Campaign-level summaries counted each distinct substitution once.

For levansucrase, the alignment contained 2,451 homologous sequences. Frequencies used the number of sequences aligned without a gap at the respective position as the denominator. Summary frequencies covered 89 distinct substitutions.

For ALDH, the P46368 A3M alignment contained 19,705 records, including the 506-residue query. Lowercase insertion characters were removed, and only sequences whose projected length matched the query were retained. Excluding the query left 19,704 homologous sequences. WT residues matched the query for all 148 substitutions across 104 positions. Frequencies reported in Results used all 19,704 homologs as the denominator, including sequences with a gap at the position; nongap coverage ranged from 256 to 18,931 sequences across the mutated positions. Figure 5G shows originality ratios. Counts for the four substitutions in the final ALDH variant are provided in Table S8.

## Supplementary Text

### Similarity to the pretraining set

Structural comparisons identified related folds among held-out targets with low sequence identity to the pretraining set. Of the 38 held-out reference sequences, 20 had a closest structural match with a query-normalized TM-score of at least 0.5. Sequence identities for these matches ranged from 1% to 25% (Fig. S1B).

### Zero-shot fitness prediction

Across the 19 held-out assays, ProTEUS ranked third by NDCG and second by top-10% recall (Fig. S2A). In the 13 single-mutant assays, ProTEUS had the highest mean Spearman correlation, whereas its NDCG and top-10% recall were slightly lower than those of the four-score MULTI-evolve rank-fusion comparator (Fig. S2B).

### Simulated evolution across datasets and experimental budgets

ProTEUS achieved higher Precision@10 than EVOLVEpro in 10, 9 and 10 of the 14 datasets at batch sizes of 30, 50 and 100 variants per round, respectively (Fig. S3).

With 30 variants selected per round, ProTEUS exceeded the mean EVOLVEpro result in 10 of 14 datasets by round 5. ProTEUS reached its final best fitness in round 1 for UBE2I, whereas GB1 and TrpB improved subsequently. EVOLVEpro recovered higher final best fitness for CALM1, GAL4, HRAS and GFP.

### In silico comparison of nomination strategies

The five nomination strategies differed in their use of zero-shot ranking, updated fitness predictions and fitness ranks derived from all available experimental measurements (Fig. S4; Table S4).

Strategy D achieved the highest across-dataset mean normalized cumulative DMS score in each evaluated round. By round 5, the mean was 0.886, compared with 0.866 for the updated predictor alone and 0.804 for the two-component ranking. The contributing datasets differed among rounds, with counts specified in the Fig. S4 legend.

### Exploratory analysis of laboratory activities

Among final-round levansucrase variants, the five-neighbor mean absolute activity difference was 3.524, compared with 3.030 for Hamming-distance-matched references (two-sided permutation $P = 0.176$). Prediction of round 2 from rounds 0–1 gave MAEs of 2.007 for embedding ridge regression and 2.992 for embedding random forest, compared with 2.007 for the training mean. Thus, these activity-based analyses did not show an advantage for the embedding readouts in the final round.

Prediction from the preceding ALDH batch yielded a Spearman correlation of 0.26 (95% confidence interval, −0.10 to 0.59; two-sided permutation $P = 0.158$). MAE was 1.60 for the embedding-neighbor predictor, compared with 1.71 for the training-mean comparator and 1.80 for the training-median comparator (Table S7). The estimated MAE reduction relative to the training mean was 0.12 (95% confidence interval, −0.16 to 0.39). Hamming-distance ties gave every training variant equal weight, producing the same predictions as the training-mean comparator. The intervals included zero correlation and zero MAE reduction, leaving prediction across these batches unresolved for this nearest-neighbor readout.

# Supplementary Figure

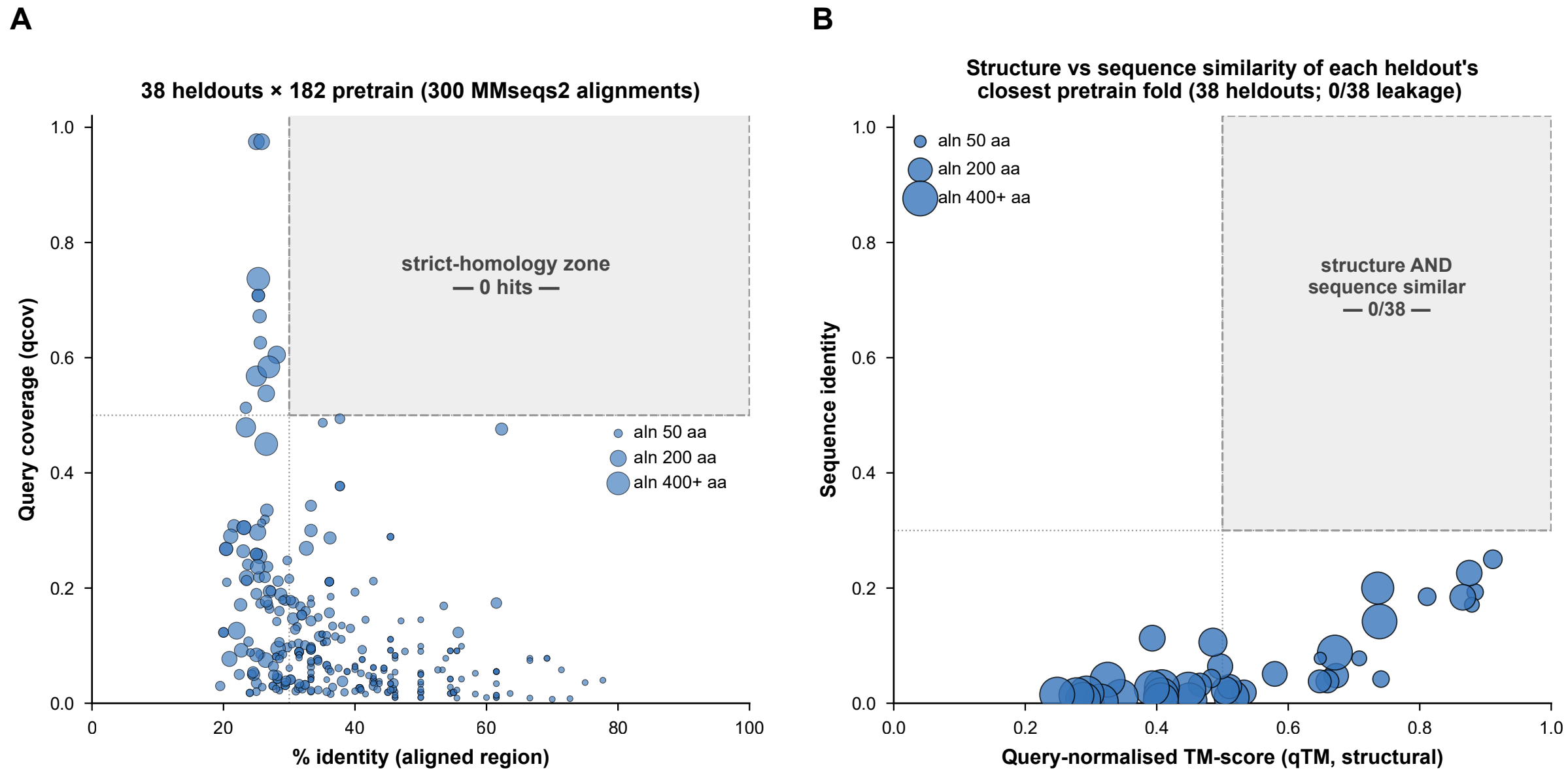

**Fig. S1. Sequence and structural similarity between held-out targets and the pretraining set**

(A) Sequence identity and query coverage for 300 MMseqs2 alignments obtained by searching 38 held-out reference sequences against sequences associated with the 182 pretraining assays. The alignments involve 33 held-out reference sequences; five reference sequences had no reported alignment. Each point represents one alignment. Sequence identity is calculated over the aligned region, and query coverage is the fraction of the query covered by the alignment. Point size indicates alignment length. Dashed guides mark 30% sequence identity and 50% query coverage.

(B) Sequence identity and query-normalized TM-score for the closest structural match of each of the 38 held-out reference sequences to the pretraining structures. The match with the highest query-normalized TM-score was retained for each query after Foldseek/TMalign comparison. Point size indicates alignment length. Dashed guides mark a TM-score of 0.5 and sequence identity of 0.3. Each point represents one held-out reference sequence.

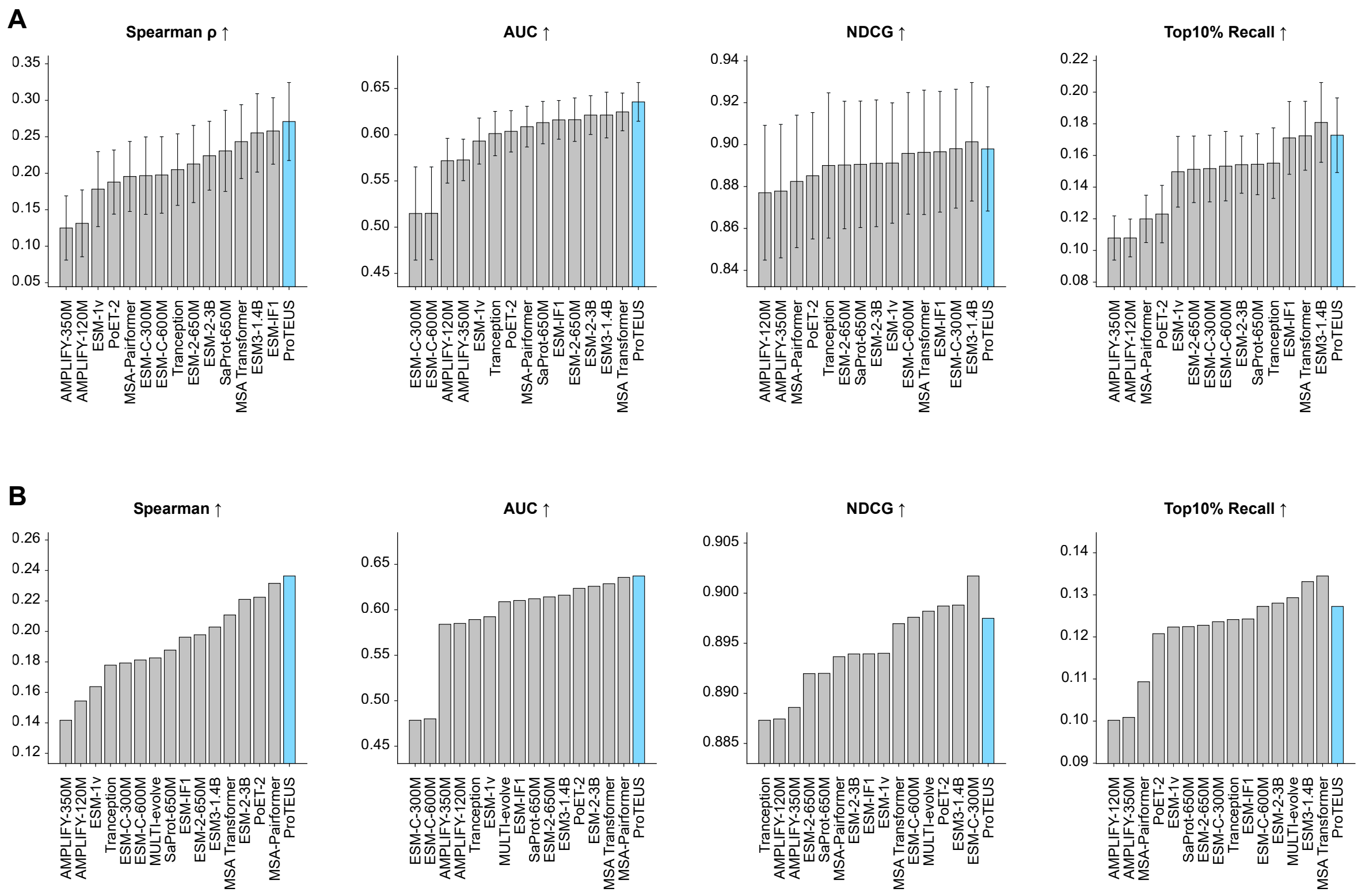

**Fig. S2. Extended zero-shot fitness prediction benchmarks**

(A) Zero-shot predictions from 15 methods across 19 held-out DMS assays, evaluated by Spearman correlation, area under the receiver operating characteristic curve (AUROC), full-list NDCG, and top-10% recall. Bars and error bars show means and SEM across assays. Missing Spearman values are assigned zero before aggregation over all 19 assays. Other metrics are averaged over available assay-level values. AUROC is available for 15 assays for most methods, 14 for Tranception, and 19 for the two ESM-C models. NDCG and recall use 19 assays except for Tranception, which uses 17.

(B) The same four metrics for 16 methods across the 13 single-mutant assays, including MULTI-evolve rank fusion. Bars show means without error bars. Available values are averaged for each metric. Spearman correlation, NDCG, and recall use 13 assays except for Tranception, which uses 12. AUROC uses 10 assays for most methods and 13 for the two ESM-C models. AUROC in both panels uses a threshold of zero on the DMS score supplied to the metric calculation. NDCG is evaluated over the full ranked list rather than a restricted top fraction.

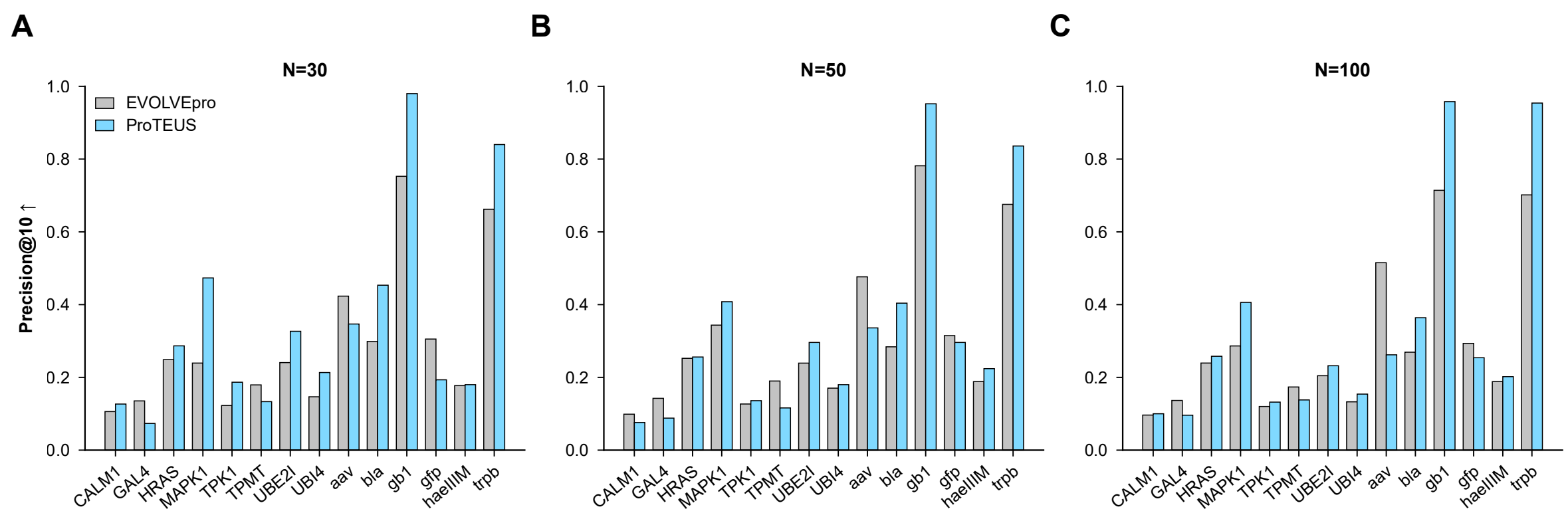

**Fig. S3. Variant-selection performance across datasets and batch sizes**
Precision@10 for ProTEUS and EVOLVEpro in simulations of directed evolution across 14 DMS datasets, with batch sizes of (A) 30, (B) 50 and (C) 100 variants per round. Precision@10 is the fraction of selected variants in the top 10% of the dataset by measured fitness. Bars show the result of one ProTEUS run and the mean of 10 EVOLVEpro runs for each dataset and batch size. The panel comprises 10 single-mutant datasets together with AAV, GB1, GFP, and TrpB.

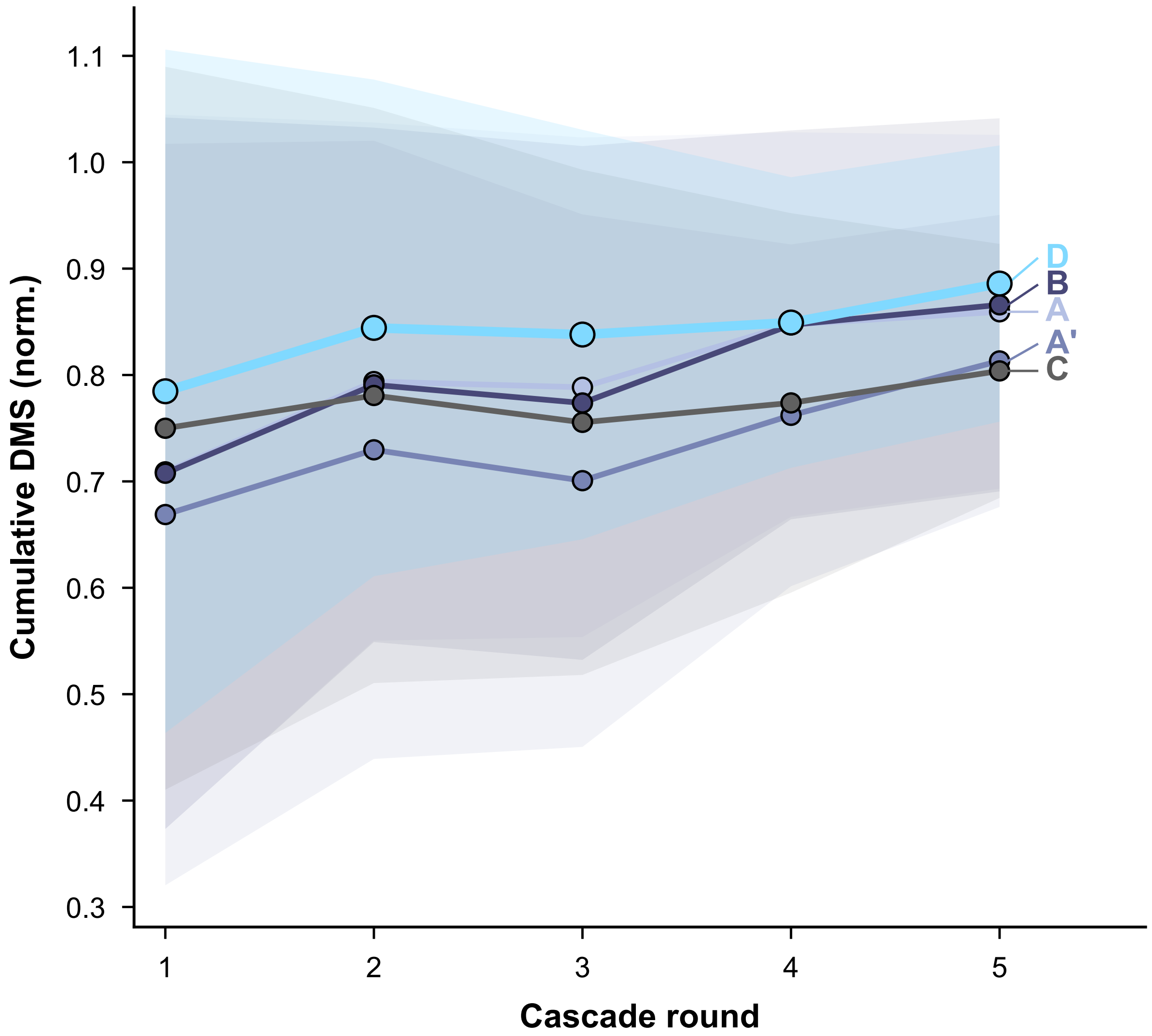

**Fig. S4. In silico comparison of nomination strategies**

Five strategies used fixed ProTEUS representations and a round-wise updated random-forest predictor (Table S4). A and A′ restricted random-forest ranking to the top 2,000 candidates or top 30% by zero-shot rank, respectively. B used random-forest ranks alone; C combined zero-shot and random-forest ranks; D additionally incorporated ranks based on the highest measured fitness among a candidate's constituent single mutants.

Each simulation began with 30 zero-shot-selected single mutants, followed by batches of 30 untested variants. The first model-guided round admitted single and double mutants; the maximum mutation order then increased by one per round, with untested lower-order variants remaining eligible. Cumulative mean DMS scores were normalized within each dataset by the largest value across all strategies and rounds. Lines and shading show the across-dataset mean and SD. The panel comprised 13 CombinGym phenotype datasets and AAV, GB1 and TrpB, with 15, 16, 13, 11 and 11 datasets contributing to rounds 1–5, respectively.

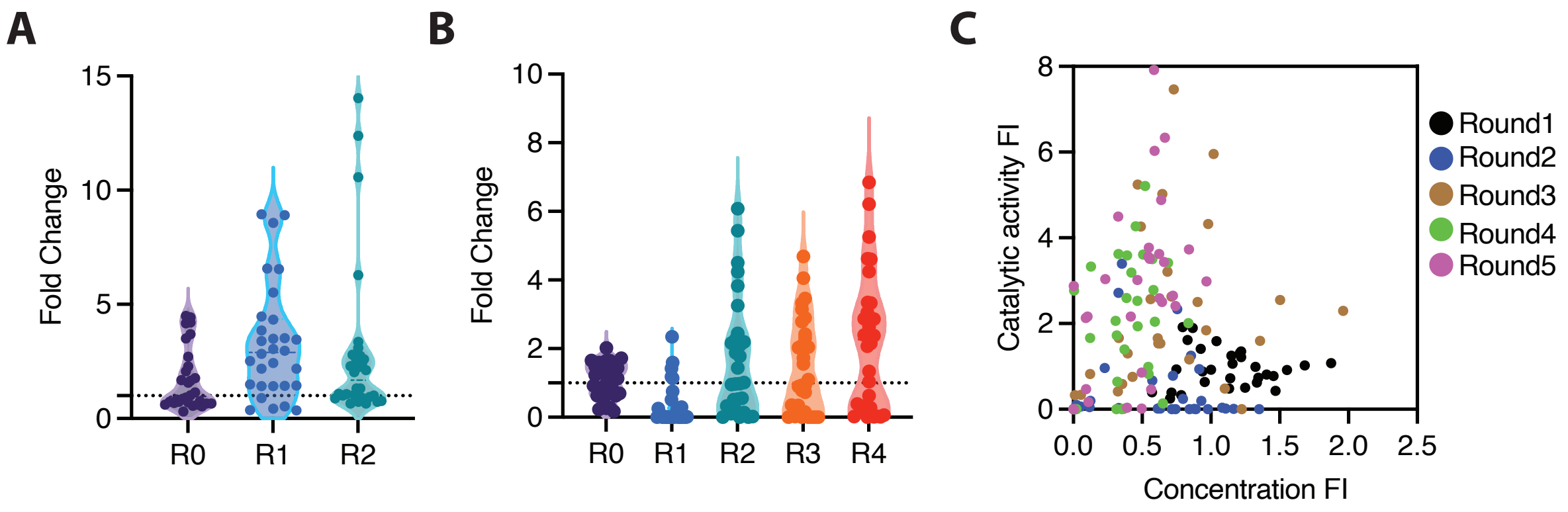

**Fig. S5. Activity distributions and enzyme abundance during laboratory evolution**

(A and B) Total protein activities (product of expression and catalytic activity) of all 90 levansucrase variants across three rounds (A) and all 150 ALDH variants across five rounds (B), expressed relative to WT. Each point represents the mean of an assayed variant (n = 30 per round); violin widths indicate the activity distributions, and dotted horizontal lines indicate WT activity (WT = 1). (C) Enzyme concentration and concentration-normalized catalytic activity of ALDH variants. Enzyme concentration was measured using an anti-His tag ELISA, and both quantities are expressed relative to WT. Colors identify experimental batches; Round1–Round5 in the legend correspond to rounds 0–4 in (B). Complete substitution combinations are listed in Tables S10 and S11 for levansucrase and ALDH, respectively.

# Supplementary Tables

### Table S1. Input modalities and encoder feature dimensions used by ProTEUS

| Modality | Input | Pretrained encoder | Feature dimension | Projected dimension |
|---|---|---|---|---|
| Sequence | Variant amino acid sequence | ESM-2 3B (12) | 2,560 | 768 |
| Evolutionary | Multiple sequence alignment | MSA Transformer (15) | 768 | 768 |
| Structural | Protein structure | ESM3 (21) | 1,536 | 768 |

Feature dimensions are reported per residue. Modality-specific adapters map each feature vector through a 64-dimensional bottleneck to the shared 768-dimensional space used by MutationFormer.

### Table S2. Learning procedures used by ProTEUS

| Procedure | Measurements used | Parameters updated |
|---|---|---|
| DMS pretraining | Training variants from 182 assays | Adapters, modality embeddings, register token, MutationFormer and neural fitness head |
| Zero-shot ranking | No fitness measurements from the evaluation target | None; sequence/MSA likelihood changes and learned modality weights are used for scoring |
| Target-specific neural fine-tuning | Target-assay training measurements from cross-validation folds or lower-order variants, depending on the benchmark | The same trainable components as pretraining; source encoders remain fixed |
| Iterative evolution | All measurements accumulated for the target so far | Target-specific random forest; ProTEUS representations remain fixed |

### Table S3. Geometric and stereochemical constraints

| Constraint family | Applied restriction |
|---|---|
| Steric separation | Nonbonded atoms in distinct chains separated by at least 0.725 times the sum of their van der Waals radii |
| Tetrahedral chirality | Improper-torsion restrictions consistent with assigned R/S stereochemistry |
| Double-bond stereochemistry | Torsional restrictions consistent with assigned E/Z configuration |
| Carbon double-bond planarity | Improper-torsion restriction enforcing local planarity |
| Ligand internal distances | RDKit-derived lower and upper bounds, scaled by 0.8 and 1.2, respectively |
| Symmetric-chain separation | Centroids of qualifying symmetric chains separated by at least 1 Å |
| Inter-chain covalent bonds | Bonded atoms in separate chains no farther apart than 2 Å |

Projection applies 20 Gauss–Seidel sweeps after coordinate prediction. These checks evaluate the specified geometry and stereochemistry.

**Table S4. Nomination strategies evaluated in simulation benchmarks**

| Strategy | Candidate ranking |
|---|---|
| A | Retain the top 2,000 zero-shot candidates; rank them by the updated random forest |
| A′ | Retain the top 30% by zero-shot score; rank them by the updated random forest |
| B | Rank all eligible candidates by the updated random forest |
| C | Combine within-library zero-shot and random-forest ranks |
| D (ProTEUS) | Combine zero-shot and random-forest ranks with fitness ranks derived from all available experimental measurements |

In the simulation benchmark, strategy D derived this ranking component from the highest available measured fitness among a candidate's constituent single mutants. Previously tested variants are excluded from every strategy's eligible set.

**Table S5. Laboratory evolution campaigns**

| Protein / round | Tested depth and background | Best variant in the batch | Activity relative to WT |
|---|---|---|---|
| Levansucrase / 0 | Single mutants from WT | Q34N | 4.50 |
| Levansucrase / 1 | Double mutants from Q34N | Q34N:G240W | 8.94 |
| Levansucrase / 2 | 12 triple mutants from Q34N:S359F; 18 from Q34N:G240W | G23A:Q34N:G240W | 14.03 |
| ALDH / 0 | Single mutants from WT | L358I | 2.02 |
| ALDH / 1 | Double mutants from L358I | A116K:L358I | 2.35 |
| ALDH / 2 | Triple mutants from A116K:L358I | A116K:H332N:L358I | 6.08 |
| ALDH / 3 | Quadruple mutants from A116K:H332N:L358I | R49K:A116K:H332N:L358I | 4.69 |
| ALDH / 4 | Quadruple mutants from A116K:H332N:L358I | S103V:A116K:H332N:L358I | 6.85 |

Each round tested 30 variants. Values are screening activities relative to WT and are distinct from the normalized slopes measured using purified ALDH. The levansucrase double mutant Q34N:S359F had activity 8.91 times WT and was retained as an alternative background for the final round.

**Table S6. ALDH local activity association across neighborhood sizes**

| Neighbors | n | Neighbor MAD | Matched MAD | Reduction (%) | Two-sided P |
|---|---|---|---|---|---|
| 3 | 60 | 1.7465 | 1.8716 | 6.69 | 0.2766 |
| 5 (primary) | 60 | 1.6900 | 1.8792 | 10.07 | 0.0456 |
| 10 | 60 | 1.7805 | 1.8934 | 5.96 | 0.0940 |

MAD denotes mean absolute activity difference, with WT activity set to 1; n denotes the number of variants. Reference sets match amino acid Hamming distance within each batch. Two-sided *P* values use 19,999 within-batch activity-label permutations without multiplicity adjustment. The five-neighbor analysis was primary; three- and ten-neighbor analyses assessed sensitivity. Relative reductions were calculated before rounding.

**Table S7. Exploratory prediction from the preceding ALDH batch**

| Predictor | Spearman correlation | Correlation 95% CI | MAE | Two-sided P |
|---|---|---|---|---|
| Embedding 5-nearest-neighbor | 0.265 | -0.10 to 0.59 | 1.5958 | 0.158 |
| Training-set mean | NA | NA | 1.7119 | NA |
| Training-set median | NA | NA | 1.7993 | NA |
| Hamming-distance neighbors | NA | NA | 1.7119 | NA |

All predictors used the same 30 training variants and 30 test variants. MAE is expressed in WT-normalized activity units. NA denotes an undefined correlation for a constant predictor. The correlation interval is a percentile bootstrap confidence interval based on 10,000 test-variant resamples. Hamming-distance ties gave all training variants equal weight in every test prediction. The MAE reduction relative to the training mean was 0.1161 (95% paired bootstrap CI −0.1566 to 0.3913); the Hamming-distance comparator gave the same result.

**Table S8. Alignment counts for the substitutions in the final ALDH variant**

| Substitution | Substituted-residue count | Homolog denominator | Frequency (%) | Originality ratio |
|---|---|---|---|---|
| S103V | 1,068 | 19,704 | 5.42 | 0.107132 |
| A116K | 104 | 19,704 | 0.53 | 0.012298 |
| H332N | 5,460 | 19,704 | 27.71 | 0.730728 |
| L358I | 7,579 | 19,704 | 38.46 | 1.000000 |

Absolute frequencies use all 19,704 retained homologs, including sequences with a gap at the relevant position. Originality ratios divide the substituted-residue count by the highest amino acid count at the same position.

**Table S9. Levansucrase representation geometry within parental backgrounds**

| Parent | n | Pairs at Hamming 2 | Cosine distance (min–max) | PCs for 90% variance Embedding / one-hot |
|---|---|---|---|---|
| Q34N:G240W | 18 | 149 | 0.00151–0.03240 | 6 / 14 |
| Q34N:S359F | 12 | 66 | 0.00310–0.03195 | 5 / 10 |

All variants were triple mutants. n denotes the number of variants in each parental background. Cosine-distance ranges include all within-background pairs differing at two amino acid positions. Principal components (PCs) were calculated separately within each background using all variants.

**Table S10. Experimentally tested levansucrase variants**

Complete substitution combinations for the 90 assayed variants in Fig. S5A, grouped by experimental round (30 variants per round). Round 0 denotes the initial zero-shot batch. Substitutions are specified relative to WT.

| Round | Variant |
|---|---|
| 0 | F169L |
| 0 | M315I |
| 0 | N207K |
| 0 | H256K |
| 0 | F169V |
| 0 | H186N |
| 0 | K254N |
| 0 | F169I |
| 0 | M315L |
| 0 | H413Y |
| 0 | F169M |
| 0 | Q196T |
| 0 | H186D |
| 0 | Q34S |
| 0 | N207T |
| 0 | P36D |
| 0 | N382D |
| 0 | Q292K |
| 0 | S154P |
| 0 | F15L |
| 0 | M400N |
| 0 | I316V |
| 0 | A333T |
| 0 | K454D |
| 0 | H101K |
| 0 | K454N |
| 0 | Q34N |
| 0 | N207S |
| 0 | M400S |
| 0 | I374V |
| 1 | Q34N:I451E |
| 1 | Q34N:I451H |
| 1 | Q34N:K59P |

| Round | Variant |
|---|---|
| 1 | A21N:Q34N |
| 1 | Q34N:G453W |
| 1 | Q34N:K473E |
| 1 | Q34N:T456F |
| 1 | Q34N:P445Y |
| 1 | Q34N:S239W |
| 1 | Q34N:L449E |
| 1 | Q34N:R360F |
| 1 | Q34N:T267P |
| 1 | Q34N:S359F |
| 1 | Q34N:G351Y |
| 1 | Q34N:D269P |
| 1 | Q34N:T456I |
| 1 | Q34N:G240W |
| 1 | Q34N:G453Y |
| 1 | Q34N:P390F |
| 1 | Q34N:S441Y |
| 1 | Q34N:Y258P |
| 1 | Q34N:K473Y |
| 1 | Q34N:A415F |
| 1 | Q34N:Y322C |
| 1 | Q34N:V199P |
| 1 | Q34N:L449Q |
| 1 | Q34N:L384F |
| 1 | M1R:Q34N |
| 1 | A21K:Q34N |
| 1 | Q34N:T394F |
| 2 | Q34N:G222Q:G240W |
| 2 | Q34N:G240W:G453H |
| 2 | Q34N:S359F:N423Q |
| 2 | Q34N:G240W:G421D |
| 2 | Q34N:G184E:G240W |
| 2 | I3G:Q34N:G240W |
| 2 | G23A:Q34N:G240W |
| 2 | Q34N:G222W:G240W |
| 2 | Q34N:G240W:I369H |
| 2 | Q34N:G240W:T338S |

| Round | Variant |
| --- | --- |
| 2 | Q34N:I217D:G240W |
| 2 | Q34N:G240W:W353S |
| 2 | Q34N:G240W:I369C |
| 2 | Q34N:I217P:G240W |
| 2 | Q34N:G240W:Y436E |
| 2 | Q34N:M328L:S359F |
| 2 | Q34N:V228L:S359F |
| 2 | Q34N:S359F:N422K |
| 2 | Q34N:S359F:G421K |
| 2 | Q34N:L324P:S359F |
| 2 | Q34N:Y103K:S359F |
| 2 | Q34N:S359F:I369H |
| 2 | Q34N:G240W:D403S |
| 2 | Q34N:I217Q:S359F |
| 2 | Q34N:S359F:V471Q |
| 2 | Q34N:G240W:T456G |
| 2 | Q34N:S359F:V424K |
| 2 | Q34N:G240W:M364E |
| 2 | Q34N:G240W:G453D |
| 2 | Q34N:L355P:S359F |

**Table S11. Experimentally tested ALDH variants**

Complete substitution combinations for the 150 assayed variants in Fig. S5B, grouped by experimental round (30 variants per round). Round 0 denotes the initial zero-shot batch. Substitutions are specified relative to WT.

| Round | Variant |
|---|---|
| 0 | N373A |
| 0 | C366V |
| 0 | S38N |
| 0 | S234A |
| 0 | K386Q |
| 0 | I104W |
| 0 | Q56A |
| 0 | I216V |
| 0 | K96E |
| 0 | L358I |
| 0 | K96D |
| 0 | M3A |
| 0 | H332N |
| 0 | H332D |
| 0 | T114L |
| 0 | M3T |
| 0 | I147V |
| 0 | T77P |
| 0 | F443Y |
| 0 | N373V |
| 0 | G337D |
| 0 | C366L |
| 0 | T172A |
| 0 | Q55A |
| 0 | G337E |
| 0 | V110I |
| 0 | K323A |
| 0 | T77A |
| 0 | A461L |
| 0 | T40S |
| 1 | L358I:S499A |
| 1 | T77M:L358I |
| 1 | L358I:M484Q |

| Round | Variant |
|---|---|
| 1 | L358I:T491A |
| 1 | K266N:L358I |
| 1 | L358I:Y471D |
| 1 | L358I:W455Q |
| 1 | A252W:L358I |
| 1 | L358I:T491K |
| 1 | T113E:L358I |
| 1 | L358I:Y488P |
| 1 | L358I:A466N |
| 1 | L358I:G470T |
| 1 | L358I:V496M |
| 1 | S267E:L358I |
| 1 | L358I:Q490P |
| 1 | A252Y:L358I |
| 1 | L358I:Q490Y |
| 1 | L358I:A440W |
| 1 | L358I:V389F |
| 1 | T77W:L358I |
| 1 | P29N:L358I |
| 1 | L358I:V374Y |
| 1 | L358I:T491S |
| 1 | T77F:L358I |
| 1 | L358I:V410H |
| 1 | S76V:L358I |
| 1 | L358I:G470K |
| 1 | A116K:L358I |
| 1 | H150W:L358I |
| 2 | A116K:H332N:L358I |
| 2 | T78N:A116K:L358I |
| 2 | L9F:A116K:L358I |
| 2 | A116K:E288K:L358I |
| 2 | E5Q:A116K:L358I |
| 2 | T77N:A116K:L358I |
| 2 | S12Y:A116K:L358I |
| 2 | T77D:A116K:L358I |
| 2 | A64L:A116K:L358I |
| 2 | P14W:A116K:L358I |

| Round | Variant |
|---|---|
| 2 | P14F:A116K:L358I |
| 2 | L9W:A116K:L358I |
| 2 | A116K:D357N:L358I |
| 2 | A4E:A116K:L358I |
| 2 | A68L:A116K:L358I |
| 2 | T77Y:A116K:L358I |
| 2 | N2Q:A116K:L358I |
| 2 | A116K:L276I:L358I |
| 2 | T77E:A116K:L358I |
| 2 | R90P:A116K:L358I |
| 2 | A116K:V259D:L358I |
| 2 | A116K:T338M:L358I |
| 2 | A116K:T336N:L358I |
| 2 | A116K:L261N:L358I |
| 2 | R74N:A116K:L358I |
| 2 | A81I:A116K:L358I |
| 2 | A116K:R320L:L358I |
| 2 | Y15N:A116K:L358I |
| 2 | A64N:A116K:L358I |
| 2 | A116K:I257L:L358I |
| 3 | R49K:A116K:H332N:L358I |
| 3 | A116K:R316P:H332N:L358I |
| 3 | E20K:A116K:H332N:L358I |
| 3 | A4Y:A116K:H332N:L358I |
| 3 | Q18E:A116K:H332N:L358I |
| 3 | A116K:H332N:L358I:M483E |
| 3 | A116K:P226E:H332N:L358I |
| 3 | A116K:H332N:L358I:L420D |
| 3 | I6W:A116K:H332N:L358I |
| 3 | G44T:A116K:H332N:L358I |
| 3 | A116K:D315S:H332N:L358I |
| 3 | K16E:A116K:H332N:L358I |
| 3 | A116K:H332N:L358I:R478V |
| 3 | L84C:A116K:H332N:L358I |
| 3 | A116K:H332N:L358I:A364Y |
| 3 | Q8Y:A116K:H332N:L358I |
| 3 | A116K:H332N:L358I:G477F |

| Round | Variant |
|---|---|
| 3 | R80T:A116K:H332N:L358I |
| 3 | A116K:Q310E:H332N:L358I |
| 3 | A116K:H332N:L358I:R397K |
| 3 | A116K:H332N:L358I:S499E |
| 3 | A116K:R316Y:H332N:L358I |
| 3 | I42N:A116K:H332N:L358I |
| 3 | A116K:A129V:H332N:L358I |
| 3 | A116K:R126L:H332N:L358I |
| 3 | E5Y:A116K:H332N:L358I |
| 3 | A116K:I204F:H332N:L358I |
| 3 | A116K:Q310G:H332N:L358I |
| 3 | A116K:E319F:H332N:L358I |
| 3 | R52K:A116K:H332N:L358I |
| 4 | L9F:A116K:H332N:L358I |
| 4 | R90S:A116K:H332N:L358I |
| 4 | R90A:A116K:H332N:L358I |
| 4 | I104V:A116K:H332N:L358I |
| 4 | Q8T:A116K:H332N:L358I |
| 4 | G10V:A116K:H332N:L358I |
| 4 | S12L:A116K:H332N:L358I |
| 4 | A7V:A116K:H332N:L358I |
| 4 | Q8V:A116K:H332N:L358I |
| 4 | T78Y:A116K:H332N:L358I |
| 4 | I6H:A116K:H332N:L358I |
| 4 | Q8F:A116K:H332N:L358I |
| 4 | G10K:A116K:H332N:L358I |
| 4 | A116K:H332N:L358I:M445F |
| 4 | W27Y:A116K:H332N:L358I |
| 4 | S103V:A116K:H332N:L358I |
| 4 | R49D:A116K:H332N:L358I |
| 4 | W72Y:A116K:H332N:L358I |
| 4 | A116K:H332N:D357F:L358I |
| 4 | R52Y:A116K:H332N:L358I |
| 4 | W27D:A116K:H332N:L358I |
| 4 | R90I:A116K:H332N:L358I |
| 4 | R80K:A116K:H332N:L358I |
| 4 | A116K:H332N:A342Y:L358I |

| Round | Variant |
|---|---|
| 4 | V50H:A116K:H332N:L358I |
| 4 | I91E:A116K:H332N:L358I |
| 4 | D105W:A116K:H332N:L358I |
| 4 | A116K:R246L:H332N:L358I |
| 4 | R52N:A116K:H332N:L358I |
| 4 | T48W:A116K:H332N:L358I |